\documentclass[preprint]{aastex}	 
\usepackage{epsfig,emulateapj5}

\newcommand{\km}{${\rm km\,s}^{-1}$}
\newcommand{\fuse}{{\em FUSE}}

\slugcomment{Accepted Version}
\shortauthors{Lehner et al.}
\shorttitle{{\em FUSE}\ Survey of the LISM}
\begin{document}

\title{{\em Far Ultraviolet Spectroscopic Explorer}\ 
Survey of the Local Interstellar Medium  within 200 Parsec}

\author{N.\ Lehner\altaffilmark{1,2},
	E.B.\ Jenkins\altaffilmark{3},
	C.\ Gry\altaffilmark{4,5},
	H.W.\ Moos\altaffilmark{1},
	P.\ Chayer\altaffilmark{1,6},	
	S.\ Lacour\altaffilmark{1}
	}

\altaffiltext{1}{Department of Physics and Astronomy, The Johns Hopkins University,
Bloomberg Center, 3400 N. Charles Street, Baltimore, MD 21218.}
\altaffiltext{2}{Current address: Department of Astronomy, University
of Wisconsin, 475 North Charter Street, Madison, WI 53706. nl@astro.wisc.edu}
\altaffiltext{3}{Princeton University Observatory, Princeton, NJ 08544.}
\altaffiltext{4}{{\em ISO} Data Center, ESA, Villafranca del Castillo, P.O. Box 50727,
28080 Madrid, Spain.}
\altaffiltext{5}{Laboratoire d'Astronomie Spatiale, B.P. 8, F-13376 Marseille, France.}
\altaffiltext{6}{Department of Physics and Astronomy, University of Victoria, P.O. Box 3055,
Victoria, BC V8W 3P6, Canada.}

\begin{abstract}
We present {\em Far Ultraviolet Spectroscopic Explorer}\ observations of 
the interstellar gas toward
30 white dwarf and 1 subdwarf (SdO) stars. These sightlines  
probe the Local Bubble (LB) and the local interstellar medium (LISM) near the LB.  
We systematically measure the column densities for the following
species:  \ion{C}{2}, \ion{C}{2*}, \ion{C}{3}, 
\ion{N}{1}, \ion{N}{2}, \ion{O}{1}, \ion{Ar}{1}, \ion{Si}{2}, 
\ion{P}{2}, \ion{Fe}{2}, \ion{Fe}{3}, and H$_2$. 
Our survey detected only 
diffuse H$_2$ molecular clouds ($f_{\rm H_2} \la 10^{-4}$) along six sightlines. 
There is no evidence from this study that H$_2$ exists well inside the perimeter of the LB.
The kinematical temperature for H$_2$ is less than
the usual temperature observed in the local interstellar clouds, implying 
different gas phases in the LISM. 
The relative abundance ratios of \ion{Si}{2}, \ion{P}{2}, and
\ion{Fe}{2} give insight about the dust content. These ratios
vary, but are similar to the depletion patterns observed
in warm and halo diffuse clouds in more distant sightlines in the Galaxy. 
The \ion{N}{1}/\ion{O}{1} and \ion{Ar}{1}/\ion{O}{1} ratios are significantly
subsolar within the LB. Outside the LB a larger scatter is observed
from subsolar to solar. 
Because Ar and N are only weakly depleted into dust grains if at
all, the deficiencies of their neutral forms are likely due to photoionization.
The evidence for significant ionization of N (and hence 
Ar) is strengthened by the detection and measurement of \ion{N}{2}, which is
a dominant ion for this element toward many sightlines. 
\ion{C}{3} appears to be ubiquitous in the 
LISM toward our sightlines, but \ion{C}{2} remains the dominant ionization stage of C. 
The limits on \ion{Fe}{3}/\ion{Fe}{2} imply that \ion{Fe}{2} is the dominant ion. 
These observations imply that photoionization is the main ionization mechanism in the LISM
and do not support the existence of a highly ionized condition in the past. 
In view of the variations observed in the different atomic and
ionic ratios, the photoionization conditions vary significantly in the LB and the LISM.
The cooling rate in the LISM, $l_c$ (in erg\,s$^{-1}$\, (\ion{H}{1} atom)$^{-1}$),  
derived from the emission of the \ion{C}{2} 157.7 $\micron$ line 
has a mean value of $\log l_c = -25.59 \pm 0.32 $ dex, very similar to 
previous determinations. 
\end{abstract}


\section{Introduction}
Because of its proximity, the local interstellar medium (LISM) provides an
unique opportunity to study in detail the physics of the 
warm (partially ionized) interstellar gas. 
Such gas is a major component of the interstellar material 
in the Galaxy and in other galaxies. 
The overall characteristics (temperatures, densities, kinematics) 
of the LISM were determined with ground-based 
telescopes and the {\em Copernicus}, 
{\em International Ultraviolet Explorer (IUE)}, the {\em Hubble Space Telescope} ({\em HST})
satellites \citep[e.g.,][]{mcclintock75,bruhweiler82,frisch83,linsky93,lallement95,redfield02}. 
Measurements of interstellar gas after it has entered the heliosphere has  also
provided information on the interstellar medium very close to the sun (see, e.g., 
studies of the local interstellar cloud by Lallement 1998 and by Slavin \& 
Frisch 2002 and references therein).

From the ground-based observations, via mainly the \ion{Ca}{2} K and \ion{Na}{1} D absorption lines, 
very accurate absolute wavelength and very high spectral resolution 
provide precise information on the local gas dynamics and the 
complex velocity structure of nearby gas 
\citep[e.g.,][and references therein]{vallerga93,ferlet99}. 
UV resonance absorption lines are much
stronger than the available optical absorption lines (which are not in the dominant
ionization stages), and are therefore more sensitive probes of the warm, low density 
LISM. Major advances in determining the physical conditions 
have been carried-out with UV observations, 
especially in the last decade, with
the  Goddard High Resolution Spectrograph (GHRS) and the Space
Telescope Imaging Spectrograph (STIS) onboard {\em HST}\
\citep[e.g.,][]{linsky93,lallement95,lallement96,linsky96}. 

With high spectral resolution ultraviolet observations, the physical properties of 
the cloud surrounding our solar system (the local interstellar cloud, LIC)
can be estimated. The LIC is not very dense ($n_{\rm H} \sim 0.1$ cm$^{-3}$), but it is
warm ($T \sim 7000$ K) and partially ionized \citep[e.g.,][]{frisch83,lallement96,linsky96}. 
Near the LIC, there are similar clouds 
with different velocities \citep[see,][]{linsky00}. All these clouds
constitute collectively the LISM, and the clouds within roughly 100 pc 
are in a region called the Local Bubble (LB; the cavity
has a radius of about 100 pc in almost every direction, except toward one very low column density
direction ($\beta$\,CMa) where it extends to $\sim$200 pc; Sfeir et al. 1999).
\ion{Na}{1} absorption line studies by 
\citet{sfeir99} show that the boundary of the LB is delineated by a sharp
gradient in the neutral gas column density with increasing radius. The distance
of this boundary deduced from X-ray data is generally smaller by about 
30\% (see Snowden et al. 1998). In this study, we use the 
{\em Far Ultraviolet Spectroscopic Explorer} ({\fuse}) satellite to  survey sightlines 
with distances less than 200 pc. For the purposes of this study, this distance will
define the extent of the LISM. Even though the distances 
are not great in the LISM, the sightlines are expected usually to  have complex 
kinematical structure \citep[e.g.,][]{lallement95}. 

{\em EUVE}\ observations also have improved our knowledge of the 
ionization structure of the LISM \citep[see the review by][and references therein]{bowyer00}.
The {\em EUVE}\ data show that the neutral 
helium to neutral hydrogen ratio is about 0.07, which 
is somewhat less than the cosmic abundance of He
to H of 0.1 \citep{dupuis95}. Helium is therefore more ionized than
H. Also, He shows less variability in its degree of ionization
from one region to the next \citep{wolff99}. 
One possible explanation for the high fractional ionization of He in the LISM
is that there has not been enough time for the 
ionized He to recombine to an equilibrium concentration from a more
highly ionized condition in the past, perhaps from the influence of
radiation from a supernova or its shockwave less than few $10^6$ years ago
\citep{frisch96,lyu96}.
A different alternative is that the LISM is currently exposed to a strong, steady flux of 
photons with $E > 24.6$ eV (e.g., a conductive interface
between the LISM and a surrounding medium at $T \sim 10^6$ K; Slavin \& Frisch 2002,
or by recombination radiation from highly ionized but cool gases; Breitschwerdt \& Schmutzler 1994).
A way to distinguish between these possibilities is to study the lightly depleted elements
Ar, N, and O \citep{sofia98}. In particular, 
the ionization of Ar (and to a lesser extent of N) can help us to differentiate between
photoionization equilibrium and non-equilibrium cooling models.

It is possible to increase
our knowledge of the ionization structure of the LISM significantly by using data from
{\fuse}\ \citep{jenkins00,lehner02,moos02,wood02b}. For instance, \citet{jenkins00} found a deficiency of \ion{Ar}{1} 
toward 4 white dwarf (WD) stars, thus favoring  photoionization equilibrium.
In the present work, we increase the sample to 31 stars 
that are located within 200 pc. 
WDs are particularly good background objects for studying the LISM because
they are nearby, they
have relatively simple stellar continua, and they are often
nearly featureless. {\fuse}\ gives access to the wavelength interval between 905 \AA\ and 
1187 \AA, a region containing features of some important neutral species 
(\ion{N}{1}, \ion{O}{1}, and \ion{Ar}{1}), many
ionized species (\ion{C}{2}, \ion{C}{3}, \ion{N}{2}, \ion{P}{2}, \ion{Fe}{2}, 
and \ion{Fe}{3}), and H$_2$.
Surveys of \ion{O}{6} interstellar absorption and \ion{D}{1}/\ion{O}{1} ratio
will be discussed by \citet{oegerle03} and \citet{hebrard02b}, respectively. 
Although {\fuse}\ has a moderate resolution of 
$R\sim 20,000$ and, hence, only the properties of the dominant cloud or
the ``average'' properties of the clouds can be derived, the large number of
resonance lines for atoms in several ionization stages as well as molecular lines provides the unique opportunity 
to study in detail the ionization structure, dust,  and molecules
in the LISM.

\section{Observations}
\subsection{The Sample}\label{sample}
In Table~\ref{t1}, we list the positions, spectral types, distances,
and heliocentric velocities for the sample of 31 stars observed with 
{\em FUSE}.  The distribution of the targets on the sky is shown in Galactic coordinates in
Figure~\ref{oimap}.  There is a good coverage of the sky in almost each quadrant with the exception 
of the region spanned by $l = 180\degr-360\degr$ and $b \ge 0\degr$. 
Most of the WDs are from the {\fuse}\ PI Team programs
(P104 and P204) used to study D/H and \ion{O}{6} in the LISM. 
The primary criterion for selecting the stars in Table~\ref{t1} was that we would be 
able to measure the species of interest for this work (see \S~\ref{col}) with 
a sufficient degree of confidence.  

Recently, in the course of the {\fuse}\  observatory program 
``Survey of White Dwarfs from the McCook and Sion Survey'' (Z903), several other WDs were observed. 
The quality of these data are not good  enough for this study
because strong airglow emission lines are present at all wavelengths
and the signal-to-noise levels are too low (the flux of these WDs in this
program are systematically lower than in the PI team programs). We also
note that these targets do not fill in the deficiency of targets in
the quadrant at $l = 180\degr-360\degr$, $b \ge 0\degr$.

\subsection{Instrument and data reduction}
The {\fuse}\/ instrument consists of four channels: two optimized for the short
wavelengths (SiC\,1 and SiC\,2; 905--1100 \AA) and two optimized 
for longer wavelengths (LiF\,1 and LiF\,2; 1000--1187 \AA). There is,
however, overlap between the different channels, and, generally,
a transition appears in at least two different
channels. More complete descriptions of the 
design and performance of the {\fuse}\/ spectrograph are given 
by  Moos et al. (2000) and Sahnow et al. (2000).
To maintain optimal spectral resolution the individual channels 
were not co-added together. Also, measurements on the independent 
spectra could be compared with each other to obtain information about the 
effects of detector fixed pattern noise. Detailed information 
for each sightline, including whether the low resolution 
(LWRS) apertures or the medium resolution (MDRS) apertures 
were used, can be found in Table~\ref{t1}.

Standard processing with the current version of the calibration
pipeline software (version 2.0.5 and higher) was used to extract
and calibrate the spectra. The software screened the data
for valid photon events, removed burst events, corrected
for geometrical distortions, spectral motions, satellite orbital
motions, and detector background noise, and finally applied
flux and wavelength calibrations.
The extracted spectra associated with the separate exposures 
were aligned by cross-correlating the positions of
absorption lines, and then co-added. In some cases, the lack 
of strong absorption lines or the contamination of airglow lines
prevented us from determining the relative displacements, 
so we simply coadded the spectra with no shifts.
Generally, the different exposures 
were obtained successively, so no large shifts were expected and no significant loss of 
spectral resolution was observed. The co-added spectra
were finally rebinned by 4 pixels ($\sim 27$ m\AA) since the extracted data are oversampled.

Figure~\ref{spec} presents two examples of {\em FUSE} calibrated spectra of WDs. 
As can be seen in the figure, the
stellar continua are relatively simple and can be fitted
with low-order ($\le 3$) Legendre polynomials. The primary error sources
are statistical noise,  fixed-pattern
noise, and possible saturation effects in the lines. The fixed-pattern noise is 
reduced by comparing the same spectral feature in different channels, 
while saturation can 
be examined when there are several transitions of the same species with different
$f\lambda$ values (\S~\ref{col}).

\section{Spectral Features}\label{col}
\subsection{General Considerations}
The primary objective of this work is to measure the column densities of the neutral atoms, weakly ionized 
atoms and molecular hydrogen toward sightlines within $\sim$200 pc
in order to probe the physical conditions within the LISM. For such study 
{\em FUSE}\/
provides access to several important resonance lines, and
we have measured systematically the column densities, $N$, 
(or limits on the column densities) for the following species: \ion{C}{2}, \ion{C}{2*}, \ion{C}{3}, 
\ion{O}{1}, \ion{N}{1}, \ion{N}{2}, \ion{Si}{2}, \ion{P}{2},
\ion{Ar}{1}, \ion{Fe}{2}, and \ion{Fe}{3}. The column densities 
of H$_2$ were determined by using the Lyman and Werner
bands. We note that \ion{P}{3} and \ion{N}{3} also are 
present in the  {\em FUSE}\ bandpass and are potentially interesting ions. 
However, the \ion{P}{3} transition is too weak to be detected and 
its 3$\sigma$ upper limit does not constrain the ratio \ion{P}{2}/\ion{P}{3} in a meaningful way. 
\ion{N}{3} is blended with the saturated \ion{Si}{2} line and 
possibly with photospheric \ion{N}{3} making any measurements of this
line highly uncertain. However, upper limits obtained by assuming 
that \ion{N}{3} is not blended and lies on the linear part of the
curve of growth (COG) show that \ion{N}{3} is not the dominant ion stage of N
in the LISM. Corrections for saturation would increase the upper 
limit, but this effect is likely less than the contribution from 
the \ion{Si}{2} blend. Therefore, nitrogen is primarily 
\ion{N}{1} and \ion{N}{2} in the LISM.

In Table~\ref{t2}, we summarize the atomic/ionic transitions
used in this work.
To measure the column densities (see below),
we adopted wavelengths and oscillator strengths from 
the Morton (2000, private communication) atomic data compilation. 
This compilation
is similar to the \citet{morton91} compilation with only a few minor updates to the atomic
parameters for lines of interest in this study.
For the \ion{Fe}{2} lines, the new oscillator strengths derived by \citet{howk00}
were adopted. We note, however, that there is a new laboratory measurement of the $f$-value 
for \ion{Fe}{2} $\lambda$1144.9
\citep[$f = 0.083 \pm 0.006$,][]{wiese02}, which is about 30\% lower than the 
experimental value of \citet{howk00}.  However, 
\citet{wiese02} note that additional study is necessary for
the other transitions available in the \fuse\/ bandpass, rather
than a simple rescaling, as they are further down on the COG
than the \ion{Fe}{2} $\lambda$1144.9 line. As we use all the transitions 
available in the \fuse\ spectrum, we employ the $f$-values of \citet{howk00}. 

For molecular hydrogen, we use 
the wavelengths from \citet{abgrall93a,abgrall93b}.
The H$_2$ $f$-values were calculated from the emission probabilities given by those
authors.

\subsection{Photospheric Lines}
To varying degrees, WD stars have sharp metal photospheric lines that can mimic interstellar lines, 
and therefore it is important to identify and disentangle them. Some WDs can be pure
hydrogen (no photospheric metal lines) and some have only small amounts of metals in their
photosphere. Their metallicity is generally very sub-solar (less than a few $10^{-3}$
solar metallicity, $Z_\odot$), and only a few metal-rich WDs are as high as
$0.1\,Z_\odot$ (e.g., WD\,2211$-$495, WD\,0621$-$376). As a consequence, WDs have
relatively smooth continua with only a few stellar lines. 
Figure~\ref{spec}  presents spectra of two WDs, one pure hydrogen
and the other with some metals, with the interstellar lines flagged.  
Table~\ref{t1} summarizes the metallicity of the WDs.

The lines that can be affected by photospheric lines  
are \ion{C}{3} $\lambda$977, \ion{Ar}{1} $\lambda$1066, \ion{P}{2} $\lambda$1052, 
and \ion{Fe}{3} $\lambda$1122. 
\ion{C}{3} can be present in WDs with $20 \times 10^3 \la T_{\rm eff} \la 50 \times 10^3 $ K,
and peak at $\sim 30 \times 10^3$ K \citep[see][]{chayer95}. Thus, when the star is not pure hydrogen,
some contamination is expected for \ion{C}{3}.
Generally the stellar radial velocity is not 
different enough from the interstellar component to be able to distinguish 
between the two components. Photospheric 
\ion{Fe}{3} can be also present for  WDs with $T_{\rm eff} \la 40 \times 10^3 $ K.
Moreover, \ion{Fe}{3} $\lambda$1122.53 is blended with photospheric \ion{Si}{4} $\lambda$1122.49. 
For this species, we have a direct diagnostic for knowing if there is photospheric \ion{Si}{4} $\lambda$1122
or \ion{Fe}{3} $\lambda$1122 by inspecting the lines \ion{Si}{4} $\lambda$1128 and 
\ion{Fe}{3} $\lambda$1128 \citep[see Figure 2 in][]{chayer00}.  \ion{P}{2} and \ion{Ar}{1}
are not found in the photosphere of the WDs, but can be blended with 
\ion{Fe}{6}, \ion{Ni}{6} $\lambda$1152 and \ion{Si}{4} $\lambda$1066, respectively. 
\ion{Fe}{6} (and \ion{Ni}{6}) are present in significant quantities only 
at  $T_{\rm eff} \ga 60 \times 10^3 $ K  \citep{chayer95}.
Therefore for most of our sample, these blends with \ion{Fe}{6} and \ion{Ni}{6} are not much of a concern. For
\ion{Si}{4} $\lambda$1066 we look at longer wavelengths (e.g. \ion{Si}{4} $\lambda$1128)
for possible presence of this line.
We finally note that some of the \ion{N}{1}, \ion{O}{1}, \ion{Fe}{2}, and H$_2$
transitions may be blended with some stellar lines, but the (generally) large numbers
of transitions for these species allow us to detect if a specific line is contaminated.

\subsection{Circumstellar Lines}
Our survey used hot WDs, typically with temperatures ranging
from 30,000 to 70,000 K (the subdwarf SdO BD+28\degr4211 has
a temperature of 82,000 K; Sonneborn et al. 2002). Such 
hot WDs are emitting a large number of ionizing 
ultraviolet photons, creating Str\"omgren spheres with radii from 
8 upto 30 pc depending on the electron density and 
the temperature of the WD \citep{dupree83,tat99}.
\citet{bannister03} recently found direct evidences
of such circumstellar absorption features (in form of high ions such
as \ion{C}{4}) in the spectra of 4 WDs out of 11 present in 
our survey.  Therefore, 
our study of the ionization conditions in the LISM could
{\em a priori} be systematically biased because of the 
use of hot WDs. However, to probe these conditions, 
we employ mainly diagnostics from neutral species, such 
as \ion{N}{1}, \ion{O}{1}, and \ion{Ar}{1}, species 
not present in the completly ionized Str\"omgren spheres. 
We have found no correlation between the ratios of \ion{Ar}{1}/\ion{O}{1}
and \ion{N}{1}/\ion{O}{1} with the temperature of the WDs. 
While it is clear that the hot WDs participate 
to the overall ionization of the LISM \citep{vallerga98,slavin02}, 
we have no evidence that stronger EUV sources give
different results compared to weaker EUV sources in our
column density ratios.

\subsection{Geocoronal Emission Lines}
Contamination of the spectra  by geocoronal emission
can be significant for the LWRS apertures of the {\fuse}\ spectrograph.
Such emission lines are very strong at \ion{H}{1} Ly\,$\beta$, but decline
rapidly in strength for the higher Lyman lines. During the sunlit part of the orbit, 
airglow emission lines are also present for many of the same lines of
\ion{N}{1}, \ion{O}{1}, and \ion{Ar}{1} used
to measure the absorption by interstellar species. The importance of airglow emissions 
depends on many factors, including the fraction of sunlit time, 
the limb angle, the flux of the target, the aperture used, and
the length of individual exposures. For the stars considered 
in this study, the observational strategy was to 
reduce the contribution from airglow lines: (1) The targets were observed in 
a manner that maximized the night time fraction, ranging typically from 
30\% to 100\% (only WD\,1234+481 was observed with basically no
night time). (2) In order to remove small spectral shifts during an observation caused by 
thermally induced motion of the optics, an exposure is broken up into a large 
number of sub-exposures.  Correlation techniques using uncontaminated 
interstellar lines are used to remove the shifts between the spectra and they 
are then added.  To the extent the shifts are due to telescope mirror 
motions rather than grating motions, these shifts will smear out the airglow 
contribution.  In practice, impact of the airglow contamination is reduced by this 
technique. (3)  The objects studied are relatively bright with fluxes from a 
few $10^{-12}$ to $10^{-11}$ erg\,cm$^{-2}$\,s$^{-1}$\,\AA$^{-1}$. 
(In contrast the stars from the Z903 program have lower fluxes, long individual
exposure times, and are observed largely in day time.)  However, even under the
conditions mentioned above, the airglow could still be significant. Therefore, we took several 
steps discussed below to check for possible contamination. 

The widths of the airglow lines are determined by the size of the 
spectrograph apertures. For the LWRS aperture, the width 
of an airglow is typically 0.35 \AA, while it is typically 
0.045 \AA\ for the MDRS aperture. Figure~\ref{spec} illustrates 
the differences in strength and width of the airglow lines 
between the MDRS aperture (WD\,1800+685, night fraction of 0.85)
and LWRS aperture (WD\,0131$-$163, night fraction of 0.95). 
Hence with the LWRS aperture, 
the primary concern for measuring the interstellar lines 
considered here is the continuum placement, 
but for the MDRS aperture, the width is small enough
that the emission could fill the interstellar lines. Several of our
targets were observed with both the LWRS and MDRS apertures. The column 
densities generally agree within the errors, and in particular 
there was no systematic smaller column densities derived with the MDRS compared to those
derived with the LWRS.

\citet{feldman01} have reported the brightnesses of the terrestrial day airglow
lines using the {\fuse} instrument. We have examined 
airglow lines not blended with interstellar lines to estimate
if our interstellar measurements could suffer from airglow contamination. At $\lambda > 1080$  \AA, 
we use the \ion{O}{1} $^1$D--$^1$D$^0$ $\lambda$1152 line as an indicator for the
presence of \ion{N}{1} emission near 1134 \AA, since the lines have comparable
strengths. This airglow line can cause a continuum
placement problem for the measurement of \ion{P}{2}  $\lambda$1152
as shown in Figure~\ref{spec} in the spectrum of WD\,0131$-$163.
We were careful to use the \ion{N}{1} $\lambda$1134 lines only when these lines were
not contaminated by airglow. At $ 1000 < \lambda < 1080$  \AA, we 
checked the strengths of several 
\ion{O}{1} airglow lines at 1027, 1028 and 1042 \AA. These
lines have strengths that indicate the strengths of the \ion{O}{1} $\lambda$1039 and \ion{Ar}{1}
$\lambda$$\lambda$1048, 1066 airglow lines. We do not find any cases where the contamination
could be troublesome (in any case the 1$\sigma$ errors quoted here should be larger than the effects of contamination). 
Finally, at  $\lambda <1000$  \AA, where the most useful lines of \ion{N}{1}
and \ion{O}{1} are located, the airglow emissions are negligible based on \ion{O}{1} $\lambda$990. 
Generally, we are confident that the airglow lines do not seriously compromise our 
interstellar measurements in ways we do not recognize. 

\subsection{Interstellar Absorption Lines}
\subsubsection{Species with Only One or Two Transitions}
The apparent optical depth (AOD) method \citep{savage91} was used when only one or two lines
of the same atomic species were available. The absorption profiles were converted into apparent
optical depths per unit velocity, $\tau_a(v) = \ln[I_{\rm c}/I_{\rm obs}(v)]$, 
where $I_{\rm c}$, $I_{\rm obs}$ are the intensity without and with the absorption,
respectively.  $\tau_a(v)$ is related
to the apparent column density per unit velocity, $N_a(v)$ 
(cm$^{-2}$ (\km)$^{-1}$) through the relation 
$ N_a(v) = 3.768 \times 10^{14} \tau_a(v)/[f \lambda(\rm \AA)]$.
The integrated apparent column density is equivalent to the 
true integrated column density in cases where the absorption is weak ($\tau \lesssim 1$)
or the lines are resolved.
This method was also used to place a lower limit on a saturated line. 

For \ion{C}{2} and \ion{N}{2}, only a strong transition
is available, and therefore, except for a few cases, the 
column densities are lower limits. The \ion{P}{2} transition is 
weak for all of the sightlines considered. \ion{C}{2*} is also usually relatively
weak for the sightlines considered, but can be slightly blended with the Lyman
5--0 R(1) line of H$_2$. When hydrogen molecules are present along a sightline,
the contribution of both lines can still be measured because the shift between
the two lines is large enough.

\ion{Si}{2} has two transitions. The weakest one is free of blending
with other interstellar or stellar lines, but the strongest transition at
989.873 \AA\ is usually saturated and blended with 
\ion{N}{3} of interstellar or stellar origin. 

For \ion{Ar}{1},  two transitions are also
available at 1048.22 and 1066.66 \AA, with the latter being the weakest 
transition of the two. We can therefore check for the effect of saturation, except
when the WD is not pure hydrogen in which case \ion{Ar}{1} $\lambda$1066.66
is blended with the stellar \ion{Si}{4} $\lambda$1066.66. 

\subsubsection{Species with several ($>2$) transitions: \ion{N}{1}, \ion{O}{1}, \ion{Fe}{2}, and H$_2$}
A COG analysis was undertaken 
when several lines were present (\ion{N}{1}, \ion{O}{1}, \ion{Fe}{2}, and H$_2$) by
using the measured equivalent widths. The equivalent widths were measured by directly 
integrating the intensity, and their
errors were derived following the method described by 
\citet{sembach92}. A single component Gaussian (Maxwellian)
curve of growth was constructed in which the Doppler parameter $b$ and 
the column density were varied to minimize the $\chi^2$ between the observed equivalent
widths and a model curve of growth. For H$_2$, first
a COG was constructed for the set of lines within each rotational level 
($J= 0,...,3$). The resulting column densities and $b$-values were then
used as the starting point for a 
$\chi^2$ minimization of a COG that simultaneously included all of the
rotational levels with the column density for each level and a single
$b$ value as free parameters.

Depending on the quality of the data and the amount of matter along a given sightline,
we found that generally the column densities of
\ion{N}{1}, \ion{O}{1}, and \ion{Fe}{2} were well determined with a COG because they 
have several transitions with different
$f\lambda$ in the {\fuse}\/ bandpass. We found that fixed-pattern noise was not a serious 
problem, and the effects of saturation can be reasonably well understood. In Table~\ref{t3}, 
we compare the $b$-values of \ion{N}{1} and \ion{O}{1} obtained from the COG analysis,
and within the errors, they are generally similar. In isolated cases where there was no good constraint
on $b$, we used the $b$-value from one species to constrain the column density
of the other. Those cases are indicated by colons in Table~\ref{t3}. The different \ion{Fe}{2}
transitions generally lie on the linear part of the COG. 

Even though we considered seriously the saturation problem, with 
the {\fuse}\ spectral resolution, it still could be that narrow, 
unresolved components could be concealed by broader features. 
\citet{jenkins86} showed, however, that column densities remain accurate to $\sim$10\%
provided that the lines are not heavily saturated (central optical depth such as 
$\tau_0 < 2$) and $b$ and $\tau$ do not have a bimodal distribution function. 
Three sightlines (WD\,0501$+$527, WD\,1254$+$223, and WD\,1314$+$293) 
of our sample were obtained with high spectral resolution using STIS. 
\citet{redfield03}, but see also \citet{vidal98} and \citet{kruk02}, derived
the column densities of several species considered in our survey for these sightlines, and their total column
density measurements are in agreement within 1$\sigma$ with ours.  This shows that
{\fuse}\ measurements are reliable despite the moderate spectral 
resolution of the instrument.
It also shows that the oscillator strengths of the species available in the {\fuse}\ 
wavelengths are also reliable. 

\subsubsection{$3\sigma$ Upper Limits}
The $3\sigma$ upper limits for the equivalent widths
are defined as $W_{\rm min} = 3 \sigma\, \delta\lambda$, where 
$\sigma$ is the inverse of the continuum S/N ratio per resolution element $\delta \lambda = 0.1$ \AA\ FWHM.
The corresponding  $3\sigma$ upper limits on the column density, presented in Table~\ref{t4}, 
are obtained from the corresponding equivalent width limits and the assumption 
of a linear curve of growth. 

\subsection{Results from Other Sources}

Table~\ref{t1} lists the sources of the column densities when they were derived from 
other investigations.  
For the sightlines that are reported in the \apjs\ special issue on
the determination of D/H with {\em FUSE}\/ 
\citep{hebrard02a,kruk02,lehner02,lemoine02,sonneborn02,wood02}, 
we complemented the measurements with additional measurements of the column densities
for species considered here but not in those papers. In addition, 
for WD\,1631$+$781 and WD\,2004-605, the values reported here are a mean 
between our methods (AOD and COG) and the profile fitting method reported 
by \citet{hebrard02b}. We note that our values agreed
at $1-2 \sigma$ level. For WD\,1211$+$332, WD\,2247$+$583
WD\,2309$+$105, and WD\,2331$-$475, the column densities  are from \citet{oliveira02}. 
For  WD\,0004$+$330, they are from Oliveira et 
al. (private communication, 2003). They were also obtained using
several independent methods. In addition, we made
independent measurements of the column species 
for WD\,0004$+$330 and WD\,2331$-$475; our values agreed at the $1-2 \sigma$ level.

The column density results are summarized for each sightline
in Table~\ref{t4} for the atomic species and in Table~\ref{t5}
for the molecular hydrogen. All the errors (except upper limits) are $1 \sigma$.

 Table~\ref{t5} also lists the rotational excitation temperature of H$_2$
for each sightline. A single
straight line with a slope characterized by $T_{\rm ex}$
fits all the observed levels $J=0,...,3$, except toward 
WD\,0004$+$330 (see \S~\ref{h2}). No rotational level
higher than $J= 3$ was detected.

\section{General Findings}\label{prop}
\subsection{Sky-Distribution of the \ion{O}{1} Column Density}
\ion{O}{1} is an excellent tracer of \ion{H}{1} because
its ionization potential and charge exchange reactions with hydrogen
ensure that the ionization of \ion{H}{1} and \ion{O}{1} are strongly 
coupled. Only a few sightlines considered in this work have 
an accurate \ion{H}{1} column density measurement \citep[see][]{moos02}. Measuring 
\ion{H}{1} column densities is particularly complex for LISM 
sightlines because the \ion{H}{1} Lyman series transitions in the {\fuse}\/ 
wavelength range are generally on
the flat part of the COG. In addition, the interstellar 
lines are narrow enough that blending with the stellar
spectra can be a significant source of uncertainty. 
Therefore, we use \ion{O}{1} as a proxy of \ion{H}{1}, adopting the 
\ion{O}{1}/\ion{H}{1} ratio derived by \citet{oliveira02} 
(who considered one sightline plus the sightlines in Moos et al. 2002),
\ion{O}{1}/\ion{H}{1}\,$= (3.17 \pm 0.19) \times 10^{-4}$, to estimate the
amount of \ion{H}{1} along a given sightline. This ratio 
is similar to values obtained for more distant sightlines 
\citep[\ion{O}{1}/\ion{H}{1}\,$= (3.43 \pm 0.15) \times 10^{-4}$; 
\ion{O}{1}/\ion{H}{1}\,$= (4.08 \pm 0.14) \times 10^{-4}$;][respectively]{meyer97,andre02a}, 
showing that O is usually a good proxy for H in different environments of the Galaxy. 
Also O is the most abundant species after H and He, making it a general good tracer 
of neutral gas over a large range of values of column density.  The \ion{O}{1}/\ion{H}{1}
ratio may vary in regions of elevated dust and density \citep{cartledge01} or possibly
in regions where the signature of stellar processed gas is still visible \citep{hoopes02}, 
but these conditions likely do not hold for the sightlines studied here. 
Therefore, we expect \ion{O}{1} to be a reliable proxy for \ion{H}{1}  in this
study.

Figure~\ref{oimap} shows the distribution of the \ion{O}{1} column density
(and thus \ion{H}{1}) in the Galactic sky. The diameter of each circle
is inversely proportional to the WD's distance and the intensity of the shading of 
the symbol indicates the total column density along a given line of 
sight. Because the clouds are not resolved (see below), most 
of the sightlines  have column densities between 15.5 and 16.5 dex. 
Generally, the farthest sightlines result in the largest
column densities. However, there are cases where, for example 
at $ -90\degr \le b \le 0\degr$ and $300\degr \la l \la 360 \degr $, 
sightlines with very different stellar distances (stellar distances
being only an {\em upper} limit on the distances of the interstellar
clouds) have a similar column density. At $d< 100$ pc, the North Galactic
pole sightlines seem to have much lower column densities than those near the South Galactic
pole \citep[see also,][]{genova90,redfield02,welsh99}. The lowest column density 
is found near  $l \sim 190\degr$ (WD\,0549+158) which is known for its relatively
low density  \citep{gry01,redfield02}.

\subsection{Sightlines and the Local Bubble}
A typical sightline goes through the Local Interstellar Cloud (LIC),
other warm partially ionized clouds, 
the Local Bubble (LB), and, depending on the distance of the star, might even 
go beyond the LB. With very high resolution observations ($R >10^5$--$10^6$),
the different clouds along a given sightline can be resolved as 
long as they do not occur at the same velocity. However, 
with {\fuse}, a resolution of only about 20 \km\ is achieved. 
As a result, the column densities derived  in this study
reflect the total amount of matter along a given 
line of sight. 

\citet{sfeir99} have used the absorption line studies of \ion{Na}{1} D,
a tracer of cold neutral gas, to map the boundary of the LB.
The LB consists of an interstellar cavity of low density 
neutral gas with radii between 65--250 pc in the galactic plane and $\sim$100
pc in other directions which is surrounded by
a denser neutral gas boundary (also referred as the ``wall''). The
\ion{H}{1} density corresponding to the wall is about 19.3 dex 
(20 m\AA\ equivalent width for \ion{Na}{1} D). Therefore both 
the distance and total \ion{H}{1} column density are used as a 
guide on whether or not a sightline extends beyond the edge of 
the LB. 

Based on these arguments, the clouds toward the following targets should be well inside the 
LB ($\log N($\ion{H}{1}$)< 19.0 $ dex): WD\,0050$-$332, WD\,0455$-$282, WD\,0501$+$527, WD\,0549$+$158,
WD\,0621$-$376, WD\,1254$+$223, WD\,1314$+$293, WD\,1634$-$573, WD\,2111$+$498, WD\,2152$-$548,
WD\,2211$-$495, and  WD\,2331$-$475. In contrast, the following sightlines have
$19.0 \la \log N($\ion{H}{1}$)\la 19.3 $ dex and therefore must be close to the 
boundary: WD\,1211$+$332, WD\,1615$-$154,
WD\,1636$+$351, WD\,2004$-$605, and WD\,2309$+$105.
All the other sightlines have $\log N($\ion{H}{1}$)> 19.3 $ dex, though
we note that some of them have values close to the approximate boundary. 
In short, some of our sightlines are well inside the LB, some are
very close to the wall, and some are beyond. 
We note that the column densities in the LIC are of the order of $5\times10^{16}$--$2 \times 10^{18}$
cm$^{-2}$ \citep{redfield00}. Therefore material in the LIC generally constitutes only a very 
small fraction of an observed total column density. 

We emphasize again that  even though some of the sightlines reside completely 
within the LB, it is likely that several clouds 
separated by less than 20 \km\
are present along a given sightline. Therefore, with the {\fuse}\/ observations we only derive either
the properties of the dominant (larger column density) cloud or a mixture of these 
clouds properties. This is
the case, for example, for G191--B2B and GD\,246  
where 2 or 3 clouds are detected using higher spectral resolution \citep{vidal98,oliveira02}. 
This hidden component structure can subsequently complicate the 
physical interpretation of these spectra (see \S~\ref{ions} and \ref{dust}).

\subsection{Distribution of the Column Densities for the Different Species}
In Figure~\ref{comphist}, we present the column density distribution
of \ion{O}{1}, \ion{N}{1}, \ion{Ar}{1}, \ion{Si}{2}, \ion{Fe}{2}, 
and \ion{P}{2} (based only on the measurements -- not the limits -- presented in Table~\ref{t4}). 
Recently, \citet{redfield02} presented the column density distribution
of {\em individual} clouds along sightlines within 100 pc. A comparison 
with the distribution of \ion{Fe}{2} column density  in their work 
(their Figure~9) and this work clearly shows an increase in our 
sample of $\sim 1$ dex in the column density. 
Our targets generally lie at larger distances and thus our sightlines have 
intrinsically more gas along the line of sight. This difference is also due to the fact  
that our sightlines contain several clouds, which are unresolved in the 
{\fuse} data.
 
Figure~\ref{comphist} also gives some general information about the
composition of the LISM. Not surprisingly, the highest column densities
are observed in the most abundant metals, \ion{O}{1} and \ion{N}{1}.
These high column densities also indicate
that the LB, known mostly for its hot gas that emits soft X-rays, 
nevertheless still has a substantial amount of neutral gas.
\ion{Ar}{1} appears to be deficient with respect to \ion{O}{1}, suggesting
that some of the gas is partially ionized (see \S~\ref{ions}).  \ion{Si}{2} and \ion{Fe}{2}
have similar solar abundances, but their column density distributions clearly
show that \ion{Fe}{2} is deficient with respect to 
\ion{Si}{2}, suggesting that Fe is more depleted onto dust grains. 
\ion{P}{2} is much less abundant than \ion{Fe}{2} and \ion{Si}{2}, but a comparison with
the solar abundance ($\sim -2$ dex with respect to \ion{Fe}{2} and \ion{Si}{2})
implies that Fe and Si are depleted onto dust grains, and hence 
dust in the form of silicate and iron grains must be present in the LISM (see \S~\ref{dust}). 

\section{Molecular Clouds in the LISM}\label{h2}
\subsection{Previous Studies}\label{previous}

{\em Copernicus}\/ was the first satellite to probe
the H$_2$ Lyman and Werner lines in absorption \citep{spitzer73}.
\citet{spitzer74} presented H$_2$ results toward 28 
stars lying between 110 pc and 2 kpc (we corrected these distances with the 
parallaxes measured by {\em Hipparcos}). Seven 
of these stars are situated between 111 and 166 pc
and are presented on the Galactic sky in Figure~\ref{oimap}
with the symbol ``C''. After {\em Copernicus}\/ was decommissioned, CO emission was the
primary technique used to 
probe molecules in the LISM. \citet{magnani85} discovered a very
large number of CO clouds at high galactic latitude $|b| \ge 20\degr$, and based 
on statistical arguments estimated that these clouds (known as MBM clouds) are within 100 pc. 
Two of these clouds, known as  
MBM\,12 ($(l,b)=(159\degr,-34\degr)$) and MBM\,16 ($(l,b)=(172\degr,-38\degr)$) clouds were considered 
until recently to be the nearest known molecular clouds to the Sun, at a
distance of $\sim$65 pc and $60\la d\la 95 $ pc, respectively \citep{hobbs86,hobbs88}. 
However, two recent independent estimates of the distance
place the cloud MBM\,12 at 275--300 pc \citep{luhman01,anderson02}, 
i.e. at a much larger distance than initially thought. 
These recent results could imply that the MBM clouds might be farther
away. \citet{dame87} presented a CO survey of the Galaxy at  $|b| \la 30\degr$, 
but the clouds studied were estimated to be farther away, $d \ga 140$ pc. 
\citet{trapero96} also found two other molecular cloudlets in $^{12}$CO emission
in the Galactic plane within 120 pc, $(l,b)\sim(149\degr,-5\degr)$. 

Direct evidence of H$_2$ in the LISM has been made possible again with the recent launch of {\fuse}. 
Note that because H$_2$ is observed in absorption,
the distance to the star directly places a firm upper limit on the distance of
any molecular cloud. 
Recently, \citet{gry02} used {\fuse}\
data to study H$_2$ in three lines of sight passing through the Chamaeleon 
complex ($(l,b)\sim(300\degr,-15\degr)$), which has an estimated distance of 150 pc.

\subsection{Distribution of H$_2$ in the LISM}\label{distri}
Although the {\fuse}\ instrument is sensitive to H$_2$ columns down to low 
values, $\sim 10^{14}$ cm$^{-2}$, there appears to be detectable amounts of H$_2$ along 
six sightlines out of 31 in our sample (Table~\ref{t5} summarizes the results).  
The sightlines have  distances between $94 \la d \la 159$ pc  
and lie at $60\degr \la l \la 120\degr$ and $-30\degr \la b \la 40\degr$ (see Figure~\ref{oimap}).
We also examined the lower signal-to-noise data from 
the Z903 program (see \S~\ref{sample}). The only two stars in the program (at the time of this work) 
that show H$_2$ in their spectra
are WD\,0421+740 ($(l,b)=(136.13\degr,+16.81\degr), d = 257$ pc) and WD\,1725+586 
($(l,b)=(87.22\degr,+33.92\degr)$, 
$d = 114 \pm 23$).\footnote{The distance for WD\,1725+586 is from J. Dupuis (2002, private communication).
For the other sightlines mentioned in this section the distances are from \citet{holberg98} or \citet{vennes98}
unless otherwise mentioned.} 
There is, however, no H$_2$ detection 
toward 3 other WDs lying in this region of the sky (WD\,1820+580, WD\,1845+683 $[d=125$ pc], WD\,1943+500 $[d=98$ pc]). 
We also note that H$_2$ was detected toward Feige 110 at even higher latitude
\citep[$(l,b)=(74.09\degr,-59.07\degr)$, $d = 179 \pm\,^{265}_{67}$ pc,][]{friedman02}. 
These diffuse  H$_2$ clouds roughly lie in the same direction as that of some of the
CO detections. Finally, 
another star which was not used in our study of atomic and ionic column densities, WD\,0441+467
($(l,b)=(158.49\degr,+0.47\degr), d = 110$ pc), has very strong 
H$_2$ transitions with a larger number of rotational levels compared to the other sightlines. 
This cloud could be related to the two CO clouds observed by \citet[][see below]{trapero96}.

Most or all of the H$_2$ detected in this study may be close to the boundary of the 
LB or outside it.  Of the six sightlines in Table~\ref{t5}, one, WD\,1636+351, has an 
\ion{O}{1} column indicating it is close to the inside of the LB wall, 
$\log (N($\ion{H}{1}$)) = 19.2$ dex.  The others are 0.3 to 1 dex higher.  
It is also possible to compare 
detections and non-detections along neighboring sightlines.  
Some of the Copernicus sightlines \citep{spitzer74} that show H$_2$ are marked  with a ``C''
in Figure~\ref{oimap}. Four of them are close to WD\,1615$-$154 where no H$_2$ was detected.
 As shown in Figure~\ref{oimap}, 
WD\,1800+685 ($d = 159$ pc), which is listed in Table~\ref{t5},  is near WD\,1631+781 
($d = 67$ pc), which shows no detectable H$_2$.  Likewise, WD\,2011+395 
($d = 141$ pc) and WD\,2247+583 ($d = 122$ pc), also listed in Table~\ref{t5}, are close to 
WD\,2111+498 ($d = 50$ pc), which shows no H$_2$.  Although small-scale structure 
in the medium could cause these differences, it is likely that the molecular 
clouds reported here reside at distances between $\sim$60 pc and $\sim$140 pc.
Using the \ion{Na}{1} D2 contour  
map of \citet{sfeir99} (their Figure~3), toward the general direction of WD\,1615$-$154 
($l\approx 350\degr - 0\degr$)
the edge of the LB is about 90 pc, while it is about 65 pc toward the general direction
of WD\,1631+781 and WD\,2111+498 ($l\approx 90\degr - 110\degr$). 
We are forced to conclude on the basis of both the \ion{O}{1} column densities 
and distance considerations, that aside from WD\,1636+351, which 
appears to be just at the wall, there is no evidence in the present data set for 
molecular clouds within the LB. We note, however, that \citet{sfeir99} found 
dense, cold clouds toward this direction at $\sim$85 pc.

\subsection{Properties of the Observed H$_2$} 

The total H$_2$ column densities are modest ranging from $1.7 \times 10^{14}$ cm$^{-2}$
toward WD\,2011+395 to $2.4 \times 10^{15}$ cm$^{-2}$ toward WD\,2148+286. 
Toward the other sightlines in our sample, the H$_2$ amount is less than $\sim 10^{14}$ cm$^{-2}$. 
These column densities are generally smaller than the ones derived by \citet{spitzer74}
(2 sightlines have $\log N({\rm H}_2)$ between 14.5 and 15.3 dex, the others marked in 
Figure~\ref{oimap} have $\log N({\rm H}_2) \ga$ 19 dex).

The molecular fraction is defined as  $f_{\rm H_2} \equiv 2 N ({\rm H}_2)/[N$(\ion{H}{1}$) + 2 N ({\rm H}_2)] $.
To obtain the values for the \ion{H}{1} columns, we use \ion{O}{1} as a proxy, except toward
WD\,2148+286 which has a good \ion{H}{1} determination \citep{sonneborn02}. The molecular
fractions obtained range between  $5.0 \times 10^{-6}$ and $1.4 \times 10^{-4}$ and are 
comparable to values obtained for other Galactic sightlines with 
$\log [N$(\ion{H}{1}$) + 2 N ({\rm H}_2)] \la 20$ dex \citep{savage77}. 
We note that these sightlines do not have higher depletion onto 
dust grains compared to the other sightlines (see \S~\ref{dust}).

In contrast,  \citet{trapero96} found  $N ({\rm H}_2) \sim N$(\ion{H}{1}$) \sim {\rm few} \times 10^{20}$ cm$^{-2}$
toward Perseus ($(l,b)\sim(149\degr,-5\degr)$, $d\sim 120$ pc) from $^{12}$CO measurements.
The {\fuse}\ data for
WD\,0441+467, discussed in \S~\ref{distri}, indicates a large amount 
of H$_2$ probably at a level of a few times $10^{19}$--$10^{20}$ cm$^{-2}$.\footnote{This 
star was not considered for this survey because the large contamination
of molecular lines and stellar lines, but clearly a detailed analysis of this sightline would be 
very interesting.} The stars observed by {\em Copernicus}\ are also at similar distances and
also have H$_2$ column densities between $10^{19}$--$10^{20}$ cm$^{-2}$ \citep{spitzer74}.
This indicates that the LISM has a higher density in this direction at relatively close
distances. 

We find that all the rotational levels can generally be fit with a single excitation temperature, 
$T_{\rm ex} \approx 300$ K (see Table~\ref{t5}), a value that represents all of the sightlines well.  
This follows the results of  \citet{spitzer74}
where they  found if $N(J=0) \la 10^{15}$ cm$^{-2}$, a single excitation temperature
generally fits all the observed $N(J)$. The value of $T_{\rm ex}$ reported in Table~\ref{t5} 
compares favorably with the results of \citet{spitzer74}.
Note that they also found one sightline 
($\tau$\,Sco) with a similar total H$_2$ column density (14.56 dex) but with a much larger $T_{\rm ex}\sim 1013$ K.

The H$_2$ $b$-values for the 6 sightlines in our sample range from 2.8 to 5 \km. 
This is smaller than the $b$-values derived for \ion{O}{1} and \ion{N}{1} 
which are generally larger than 5 \km (see Table~\ref{t3}).  The parameter $b$ is usually defined as
$b = \sqrt{2 kT/(A m) + v^2_{\rm turb}}$, where the symbols have their usual meaning.
So if H$_2$ and \ion{O}{1} (or \ion{N}{1}) were in the same gas phase, one would 
expect to have $b($\ion{O}{1}$)< b($H$_2)$, since the mass of  H$_2$ is lower than that of O.
This is not the case, and it would seem to imply that H$_2$ is in a different phase.
However, some of the higher values for $b($\ion{O}{1}) might be explained by the presence 
of several components separated in velocity but not all detected in H$_2$.
The H$_2$ $b$-values indicate a thermal temperature less than
900--3000 K, different from the temperature of 7000 K observed usually in the 
local interstellar gas. 
The \ion{O}{1} column densities and distances of these sightlines place these clouds
near the wall of the LB or beyond. This would appear to be in agreement with 
H$_2$ tracing colder neutral gas and hence with smaller $b$-values.

All the sightlines showing H$_2$
in our survey are located in roughly the same area, with similar upper limit distances
or, as discussed in \S~\ref{distri}, ranges of distances. They also have
similar column densities, molecular fractions, excitation temperatures, and $b$-values.
These facts suggest that the gas studied
could be the same large diffuse cloud, possibly in form of an extended thin sheet. 

Combining these results imply that the H$_2$ clouds in the LISM can be in form of diffuse clouds, 
but there are  regions (not explored with our survey) where there is enough gas and dust to shield 
the interstellar radiation field, allowing
a large molecular fraction to form \citep[see,][]{trapero96}. 

\section{Ionization of the LISM}\label{ions}
\subsection{The Neutral Species}
\subsubsection{Background}

Recent studies of the neutral species in the LISM 
reveal that they could constrain the origin (photoionization
or non-equilibrium cooling) of the ionization \citep{sofia98,jenkins00}. 
The argument of \citet{sofia98} is that a deficiency of \ion{Ar}{1} 
is not caused by Ar depleting into dust grains, but rather 
its neutral form can be more easily photoionized than H. 
The photoniozation cross-section of Ar is about 10 times
that of H over a broad range of energies, so Ar can be 
more fully ionized than H in a partially ionized gas. 
While in dense regions N behaves like O, i.e. the ionization
fractions of these elements are strongly coupled with 
the ionization fraction of H via a resonant charge-exchange
reaction, in low density, partially ionized regions 
N behaves more like Ar by showing a deficiency of its neutral form. 
N also is not depleted onto dust grains \citep{meyer97}. 
On the one hand, if a region was highly ionized at some earlier epoch, 
by say a shock, and is in the process of recombining, 
then \ion{Ar}{1}, (\ion{N}{1}) and \ion{H}{1} would have about the same 
ionized fraction because the recombination coefficients
are roughly the same. On the other hand, if steady photoionization
dominates, then \ion{Ar}{1} should be deficient
from the arguments mentioned above. If the region is
low density and ionized, \ion{N}{1} will also be deficient. 

We first describe the results that the data show 
in this section and \S~\ref{ionspecies}. We will discuss more thoroughly their meaning
and their implications with new ionization calculations in  \S~\ref{model}.

Figure~\ref{comp} presents the ratios of \ion{Ar}{1}/\ion{O}{1},
\ion{N}{1}/\ion{O}{1}, and (\ion{N}{2}+\ion{N}{1})/\ion{O}{1} 
plotted versus the total column density of \ion{O}{1} and the distance
of the sightline. Both  \ion{Ar}{1} and \ion{N}{1}
show deficiencies with respect to \ion{O}{1} toward many 
of the sightlines but the scatter is large. We also plot 
in this figure the B-type star and solar abundances for 
Ar/O, and B-type star, solar, and interstellar abundances for N/O. 

The reference abundances of Ar are from \citet{keenan90} and \citet{holmgren90} 
for B-type stars, and from \citet{anders89} and \citet{meyer89}
for the solar abundance. The solar O and N abundances are
from \citet{allende01} and \citet{holweger01}, respectively. 
We note that  \citet{holweger01} also derived an O solar 
abundance, but with a higher uncertainty than \citet{allende01}. In any
case the difference between these two values is 0.05 dex and
does not change any of our conclusions.
The interstellar O and N abundances are
from \citet{meyer98} and \citet{meyer97}, respectively
(the interstellar
abundances were typically measured for more distant sightlines
with higher hydrogen column densities, $20.2 \le \log N($\ion{H}{1}$)\le 21.2$). 
For the O and N abundances in B-type stars see \citet{sofia01} and references
therein. 
Table~\ref{t6} summarizes the \ion{Ar}{1}/\ion{O}{1} and \ion{N}{1}/\ion{O}{1}
ratios measured in this study, normalized to the solar abundances for reference.

\subsubsection{Nitrogen and Oxygen}

The \ion{N}{1}/\ion{O}{1} ratio versus the distance presented in Figure~\ref{comp} 
does not show any particular trend,
probably primarily because the interstellar clouds are
not resolved and the exact distance of the clouds remains
unknown. However, when considering the sightlines
for which the clouds are well within the LB ($\log N($\ion{O}{1}$)\la 15.60$), 
\ion{N}{1}/\ion{O}{1} is systematically down to $-1.00 \pm 0.07$ dex 
(compared to the interstellar ratio in denser clouds of $-0.66$ dex
and to the solar value of $-0.76$ dex). Once
$\log N($\ion{O}{1}$)> 15.60$ dex, i.e. the gas is near or beyond the 
edge of the LB, the picture is far more complicated, with  a scatter 
of \ion{N}{1}/\ion{O}{1} between $-1.1$ dex and 
the interstellar value obtained for sightlines with much higher
column densities; \ion{N}{1}/\ion{O}{1} varies over 0.45 dex in the LISM.

The bottom panel of Figure~\ref{comp}
shows clearly that measurements or lower limits of
(\ion{N}{2}+\ion{N}{1})/\ion{O}{1} are much closer to a normal interstellar N/O ratio. 
The deficiency of N is therefore explained by ionization of this element.
In the LB, the amount of \ion{N}{2} can be as large as $\sim$3 times the
amount of \ion{N}{1}. Several limits or values of (\ion{N}{2}+\ion{N}{1})/\ion{O}{1}
are above the interstellar N/O ratio,
suggesting that O can be partially ionized in significant quantities as well. 

The last column in Table~\ref{t6} gives the ratio of
\ion{N}{1}/(\ion{N}{1}+\ion{N}{2}). 
There is a measured value of this ratio toward 5 stars. For the 
four of them which are well inside the LB ($\log N($\ion{O}{1}$)\la 15.60$) the measured ratio
implies that from 40\% to 70\% of N is ionized, while in the sightline 
located near the edge of the LB ($\log N($\ion{O}{1}$) =  15.78$) only 5\% of the N is ionized.
The upper 
limits show, depending on the direction, that at least 2\% to 
75\% of N can be ionized. Combining the upper limits and the measurements
suggests that the ionization conditions are not uniform in the LB 
(even though \ion{N}{1}/\ion{O}{1} appears fairly constant). 

\subsubsection{Argon and Oxygen}

For most sightlines,
the \ion{Ar}{1}/\ion{O}{1} ratio also shows a deficiency of 
\ion{Ar}{1} in the LB by $\sim -0.3$ dex with respect to the solar value
when $\log N($\ion{O}{1}$) < 15.60$ dex, but with a larger scatter than the 
scatter observed in \ion{N}{1}/\ion{O}{1}.
\ion{Ar}{1}/\ion{O}{1} is near solar for the WD\,1634$-$573 sightline  
with  $\log N($\ion{O}{1}$) = 15.51$ dex. 
\citet{wood02} using GHRS G230M observations suggest that there 
could be at least two  clouds along WD\,1634$-$573. The amount of \ion{N}{2} is actually larger
than the amount of \ion{N}{1} by a factor 2 toward WD\,1634$-$573, even though the
\ion{N}{1}/\ion{O}{1} is the highest in the LB sample.
It seems therefore plausible that one of these clouds is largely ionized,
while the other one is mostly neutral. This explains why \ion{N}{1}/\ion{O}{1}
and \ion{Ar}{1}/\ion{O}{1} are near solar, since these atoms exist predominantly in the neutral cloud.
In addition, the super-solar and super-interstellar ratios for (\ion{N}{2}+\ion{N}{1})/\ion{O}{1}
and the low \ion{N}{1}/(\ion{N}{2}+\ion{N}{1}) ratio indicate
a deficiency of the neutral form of O and N, and thus the presence of 
a substantially ionized cloud along this line of sight (this is also confirmed by the 
strong \ion{C}{3} line, see \S~\ref{ionspecies}). Hence, 
a near solar ratio for \ion{Ar}{1}/\ion{O}{1} 
does not necessarily imply that there is no ionized 
cloud along the sightline. 

Near the edge of the LB and beyond,
the scatter of the \ion{Ar}{1}/\ion{O}{1} becomes larger, ranging from a solar value 
to $-0.8$ dex subsolar.  In the Local Cloud within $\sim5$ pc from the Sun,
\ion{Ar}{2} is the dominant ion according to the photoionization calculation by
\citet{slavin02}. The scatter in the \ion{Ar}{1}/\ion{O}{1} ratios is large, 
either plotted with respect to the \ion{O}{1} column density or  the distance
and is larger than the scatter of \ion{N}{1}/\ion{O}{1}.

The \ion{Ar}{2} line can be found at 919.781 \AA, which is within
the {\fuse}\ spectral range. However, the data in this wavelength 
region are usually of lower quality and
this line is too weak to be detectable in the spectra
studied here. \citet{sofia98}
show that Ar was very unlikely to be depleted onto dust grains. Therefore we conclude that
the observed deficiency of \ion{Ar}{1} found for almost every sightline in our study of the
LISM is due to photoionization.  

\subsubsection{Nitrogen and Argon}
In Figure~\ref{ratio}, we compare the ratios of [\ion{Ar}{1}/\ion{O}{1}] and 
[\ion{N}{1}/\ion{O}{1}] normalized to their solar values\footnote{Hereafter, [X/Y] refers to an 
abundance ratio in logarithmic solar units, X and Y represent
the column densities of two elements: [X/Y]\,$=\log({\rm X/Y}) - \log({\rm X/Y})_\odot$.} 
but only for data with  $1\sigma$ error $\la \pm 0.30$ dex, 
i.e. only with the higher quality measurements. 
This plot confirms that \ion{N}{1} and \ion{Ar}{1} are deficient compared to their respective
solar values. With only a few exceptions, [\ion{N}{1}/\ion{O}{1}$] \ga [$\ion{Ar}{1}/\ion{O}{1}].
This result can be explained by the fact that Ar is more easily ionized 
than N because of its larger ionization cross section.
There are, however,
several data points that suggest that \ion{N}{1} could be as efficiently ionized as
\ion{Ar}{1}. Figure~3 of \citet{sofia98} 
suggests that for ionizing photons energies $E > 35$  eV, N could be more easily
ionized than Ar. But with the constraint from \ion{C}{3}/\ion{C}{2}
measured in this study and with new ionization calculations, it appears
difficult to satisfy this condition in the LISM. We will return to a discussion of [\ion{Ar}{1}/\ion{O}{1}] and 
[\ion{N}{1}/\ion{O}{1}] in the context of a photoionization model in
\S~\ref{model}.

\subsection{The Ionized Species}\label{ionspecies}
The singly ionized species measured in this work are
in the dominant ionization stage in both the neutral
and ionized gas, and therefore cannot tell us 
directly much about the ionization conditions in the LISM. 
There are, however, detections of \ion{C}{3} 
along nearly every sightline and 
one detection of \ion{Fe}{3} toward WD\,1800+685.
The production of \ion{C}{3}  by photoionization 
requires photons with energies greater than 
the \ion{C}{2} ionization threshold of 24.4 eV. At
energies above 47.9 eV, \ion{C}{3} is ionized to \ion{C}{4}.
For the production of \ion{Fe}{3}, the \ion{Fe}{2} 
threshold is 16.2 eV and the \ion{Fe}{3} threshold is 
30.7 eV. As discussed in \S~\ref{model}, \ion{C}{3},
in particular, is a good tracer of ionized gas.

Unfortunately, usually both  the \ion{C}{2}
and \ion{C}{3} lines are saturated, and only a few meaningful
limits and measurements are presented in Table~\ref{t7}. 
In this Table, we also list the values or range of values of \ion{C}{3}
after correction for blending of the interstellar line with the stellar \ion{C}{3} line. 
We modeled the synthetic spectra by using either the \ion{C}{3} excited lines at 
1175 \AA\ or using \ion{C}{4} available with {\em IUE}.
In some cases, we were only able to estimate a limit
and hence only a range of possible values.

Toward WD\,1634$-$573 the value of \ion{C}{3}/\ion{C}{2} 
is relatively high, about $-0.63 $ dex, confirming the 
presence of a largely ionized cloud along this sightline, 
possibly a \ion{H}{2} region around the star.  
For all the other sightlines
the ratio is less than or of the order of $ -1$ dex; and toward 
WD\,1314+293 this ratio is very weak, $-2.24$ dex. 
The sightlines for which we were able to measure \ion{C}{3}/\ion{C}{2} 
lie well  within the LB. However, Tables~\ref{t4} and \ref{t7} show that
\ion{C}{3} appears to be ubiquitous in the LISM toward our 
sightlines. We note, however, that toward several more nearby stars,
\ion{C}{3} was not detected \citep{redfield02}, implying that either
the ionization conditions change in more distant sightlines or 
that \ion{C}{3} might arise partly from \ion{H}{2} regions near the WDs. 
While \ion{C}{3}/\ion{C}{2}
appears to vary from sightline to sightline, unlike what we found
for \ion{N}{2}/\ion{N}{1}, the fraction of 
\ion{C}{3} with respect to \ion{C}{2} is generally small. 
The recent ionization calculations of \citet{slavin02} of the Local Cloud
within 5 pc of the sun show
that \ion{C}{3}/\ion{C}{2}$\sim -1.44$ dex (model 17, their Table 8),
in general agreement with our measurements. 
Our observations typically
probe denser clouds. Possibly a significant fraction of the 
clouds are neutral regions, where \ion{C}{2} is also dominant. These observations demonstrate
that \ion{C}{3}, while present, is not a dominant ion in the LISM. 

While we generally have a good measurement of \ion{Fe}{2}, 
\ion{Fe}{3} is either undetected or can be blended with 
photospheric lines. Only one measurement, 
$\log [N($\ion{Fe}{3}$/N($\ion{Fe}{2}$)] \sim -1.2 $ dex, was obtained
toward WD\,1800+685 which probably lies beyond the 
wall of the LB. This result and the other limits are summarized in Table~\ref{t8}. 
\citet{slavin02} found with their ionization calculations 
\ion{Fe}{3}/\ion{Fe}{2}$\sim -1.40$ dex (model 17, their Table 8), 
which, as for \ion{C}{3}, shows an agreement with our measurements
and indicates that \ion{Fe}{2} is the dominant ion in the LISM.

\subsection{Ionization Calculations and Implications}\label{model}
To obtain more quantitative insights, we developed the photoionization code used 
by \citet{sofia98} and \citet{jenkins00}. 
The detailed equations and how they are solved are 
fully described in \citet{sofia98}. But here, in all cases, the formulae included the
effects of charge exchange with neutral hydrogen and the modifications 
of the H and He ionization fractions. We modified several parameters in the
program that created Figure~2 in \citet{jenkins00} in the following ways: 
\\
(1) We replaced the hot gas interface radiation given by \citet{slavin89}
with the spectrum calculated by \citet{slavin98}.  
This new spectrum shows more details at high photon energies, and the physical
conditions for the Local Cloud seem more realistic.  The new
spectrum is about 5 times more intense than the old one.  This slightly
increases the ionization inside the cloud. We are assuming 
that more distant clouds have properties similar to those that 
we know exist for the Local Cloud. Likewise, we
assume that the radiation field due to the WDs is the same
everywhere and is that reported for the local neighborhood by \citet{vallerga98} using {\em EUVE}\ observations.
\\
(2) The upper limit for the \ion{H}{1} absorbing column densities has been
increased to somewhat above $10^{19}$ cm$^{-2}$, on the assumption that some
lines of sight could be dominated by monolithic clouds with this
column density.  This upper limit approximately corresponds  to the
median column density of \ion{O}{1} that we found in this survey.  In all
likelihood, the higher column density cases are composed of several
clouds that have less shielding than a few times $10^{19}$ cm$^{-2}$ of
neutral hydrogen. In addition, the complex non-spherical geometry
of a cloud may reduce the effective shielding to absorbing column densities
much lower than the value implied by the measurement along the sightline. 
\\
(3) We added the atomic parameters for C, P, Si, S, and Fe. In particular,
we plot explicitly the values of \ion{C}{3}/\ion{C}{2} and 
\ion{Fe}{3}/\ion{Fe}{2}.

The results are plotted in Figure~\ref{figcal} as a function 
of the absorbing column shielding in the external ionizing radiation sources.
A uniform pressure within a given cloud of $p/k = 2.5 \times 10^3$ cm$^{-3}$\,K
is assumed. 
This pressure is consistent with the pressure derived by \citet{gry01} 
toward $\epsilon$\,CMa. Reducing the pressure would give a
\ion{C}{3}/\ion{C}{2} ratio that is too high with respect to the observed values. 

Figure~\ref{figcal} illustrates how \ion{O}{1} follows \ion{H}{1} very well, while
\ion{N}{1} and \ion{Ar}{1} become progressively more deficient as 
the shielding from hydrogen decreases. 
The other elements (C, P, Si, S, and Fe) remain primarily in the singly 
ionized state, but the abundances of these ions
are elevated relative to that of neutral hydrogen.
The assumed pressure seems to give a reasonable set of values for  
\ion{C}{3}/\ion{C}{2}, but \ion{Fe}{3}/\ion{Fe}{2} appears to be
too low. Yet, we have only one measurement and this measurement
is very uncertain (see \S~\ref{ionspecies}). In contrast, \citet{slavin02}
found similar fractions for \ion{C}{3}/\ion{C}{2} and \ion{Fe}{3}/\ion{Fe}{2} in their model. 
Except for this discrepancy, these two models are in reasonable agreement. 
According to our model, the fraction of \ion{Fe}{3} is heavily dependent 
on the effect of charge exchange with neutral hydrogen. This is not
the case for \ion{C}{3}. Without taking 
into account the effect  of charge exchange with neutral hydrogen, 
the fraction of \ion{Fe}{3} is comparable to the fraction of \ion{C}{3}. 
Therefore, the amount of \ion{Fe}{3} depends on the density and 
ionizing radiation field, making \ion{Fe}{3} a less reliable tracer
of ionized gas compared to \ion{C}{3}.

In Figure~\ref{ratio}, the model predictions are compared with the measured 
ratios of [\ion{N}{1}/\ion{O}{1}] and [\ion{Ar}{1}/\ion{O}{1}]. 
There is a large
scatter in the measurements due to the scatter in the physical
properties of the different clouds. 
The model assumes the same 
physical properties for all clouds. The study reported here
and many others \citep[e.g.,][]{gry01,redfield02} demonstrate that this is not the case.
Nevertheless, it should be noted that, within 1$\sigma$ error, 
the majority of the data points agree with our model.

For about 6/22 data points in Figure~\ref{ratio}, [\ion{Ar}{1}/\ion{O}{1}] is significantly 
higher than predicted and thus is approximatively equal to [\ion{N}{1}/\ion{O}{1}].
This suggests that, for these sightlines, the ionization fraction 
of N could be as large as that of Ar. Using our model, 
we have explored several combinations of
parameters by varying the pressure and ionizing flux to see 
if we could reproduce such ionization fractions. 
First, the value of $p/k$ must be well below $10^3$ cm$^{-3}$\,K to make the
electron density high enough (and $n($\ion{H}{1}) low enough) in order to have
the charge exchange relatively inefficient between nitrogen and hydrogen. Moreover, 
the spectrum of the ionizing radiation must be altered drastically, e.g.,
with a very strong flux of photons (of unknown origin) that have
an energy $E \ga 35$ eV (see Figure~3 of Sofia \& Jenkins 1998). 
Perhaps these photons originate from the recombination of \ion{He}{2}.
When all of these extreme conditions are invoked, the ionization of N
can even be slightly higher than that of Ar. However, under those conditions \ion{C}{3}/\ion{C}{2}
becomes far too large compared to our result. So with our model, 
a higher or comparable ionization fraction
of N compared to that of Ar is difficult to explain with a simple photoionization phenomena. 
Assuming collisional ionization equilibrium in a hot plasma, \citet{shull82} calculated the
expected deficiency of the neutral forms of Ar, N, and O. According to their
values, collisional ionization of these elements would not appear to solve this problem.
While in principle O can be depleted by about 25\% \citep{moos02}, 
Ar and N are very unlikely to be depleted
by any significant amount \citep{sofia98,meyer97}. Another possibility  
would be an overabundance of Ar in certain directions. But that
would seem to be difficult to reconcile with no apparent changes in abundances
for the other elements, especially the $\alpha$-elements such as O.  

While our model might be too naive to reproduce the different changes in the physical
conditions occurring in the LISM, our observations strongly indicate
that photoionization is a major ionization contribution in the LISM. 
These results do not support the proposition that the LISM is in the process of recombining
from an highly ionized phase at some earlier epoch, in which case \ion{Ar}{1} and
\ion{N}{1} and \ion{H}{1} would be roughly equally ionized.
The photoionization conditions may not be constant in the LB and more generally in the LISM
according to the variation observed in \ion{Ar}{1}/\ion{O}{1}, 
\ion{N}{1}/\ion{O}{1}, \ion{N}{1}/(\ion{N}{1}+\ion{N}{2}), and \ion{C}{3}/\ion{C}{2}.
But this is complicated by the unresolved cloud structure. For example, toward WD\,1634$-$573, 
at least two clouds are present with one of them being substantially ionized 
(large \ion{C}{3} and \ion{N}{2} fractions; possibly an \ion{H}{2}
region near the star) and the other one mostly neutral
(near solar abundance for \ion{Ar}{1}/\ion{O}{1} and \ion{N}{1}/\ion{O}{1}). Another example is
toward WD\,1314+293, where there is a large fraction of \ion{N}{2} but a very small
fraction of \ion{C}{3}.

\section{Dust in the LISM}\label{dust}
Several studies have examined the deficiency of 
elements in the gas-phase of the Galaxy  with respect to solar 
or B-type star abundances
\citep[see the review by][and references therein]{savage96}. 
When ionization is not the cause, these deficiencies are called
depletions and are usually explained by the fact 
that certain elements are preferentially locked up
into dust grains. 
These studies show a general progression of increasingly 
severe depletion from warm halo clouds, to warm disk clouds, to colder
disk clouds. Below, we will compare these observed depletion patterns with
studies of other parts of the Galaxy. 
\ion{S}{2} (or \ion{Zn}{2}) or \ion{H}{1} are often used because the former two
are lightly or not depleted onto dust grains and the latter gives 
the absolute gas-phase abundance directly (when ionization is not important). 
\ion{S}{2} and \ion{Zn}{2} are not available in the {\fuse}\/ wavelength range, 
and as discussed in the \S~\ref{ions}, the LB and LISM are partially ionized, 
therefore using \ion{O}{1} as a proxy for H is not suitable. 
However, we show in \S~\ref{model}
that as long as conditions are not extreme, the other elements
(C, Si, P, and Fe) appear to remain primarily in the singly ionized state.
This is also confirmed by the photoionization calculations of \citet{slavin02}.
Therefore the abundance of \ion{C}{2}, \ion{Si}{2}, \ion{P}{2}, and \ion{Fe}{2}
can be compared directly in order to measure the depletion onto grains. 
We note that this method of sensing the existence of
dust relies on the underlying assumption that the
reference abundances of the elements  apply to the LISM.
Direct detection of dust grains within the solar system were made 
possible using the {\em Ulysses}\ and {\em Galileo}\ satellites
and are discussed by \citet{frisch99}.

Previous studies show that P is only lightly depleted onto dust grains 
especially in warm gas \citep[e.g.,][]{welty99}. However, \ion{P}{2} 
is 2 or 3 orders of magnitude less 
abundant than the other species, so that only a few measurements were 
possible in our survey. On the other hand, C is very abundant 
and is lightly depleted. Because of this and the fact that
C is so abundant and the only available 
transition of \ion{C}{2} is so strong, a measurement of the column density of \ion{C}{2}
was possible toward only a few sightlines with low columns 
(the high velocity components have all low column densities and 
are discussed in \S~\ref{hvc}). \ion{Si}{2} and \ion{Fe}{2} are both depleted 
into dust grains, and they are easily measured in our survey.

Evidence for dust grains in the LB and beyond is shown
in Figure~\ref{compsifep} where the logarithmic ratios of 
$N($\ion{Fe}{2}$)/N($\ion{Si}{2}) and $N($\ion{P}{2}$)/N($\ion{Si}{2})
are shown. Even though P is the least depleted amongst these three elements
and would then be a better element for reference, we chose Si because 
there are many more measurements in common with Fe. 
The solar (meteoric) values of these ratios are 
$-0.04$ and $-1.98$ dex for Fe/Si and P/Si, respectively \citep{anders89}.
Except for one value that could be compatible with 
a solar value within the error bars (WD\,2011+395), 
[\ion{Fe}{2}/\ion{Si}{2}] is depleted from about $-0.3$ to $-0.8$ dex,
typical of what is observed in ``cold-warm'' and ``halo'' clouds
\citep[][and references therein]{savage96}. Note that
for Fe/Si it is not possible to differentiate between 
a cold and a warm depletion pattern, since it is depleted in
both types of gas by $\sim -1$ dex. 
Figure~\ref{compsifep} shows that
there is no difference between the 
interstellar clouds within the LB,
near the wall, or beyond. In particular the range of values 
over which [\ion{Fe}{2}/\ion{Si}{2}] varies
is similar.

[\ion{P}{2}/\ion{Si}{2}] varies generally from a solar value ($-2.0$ dex) to $-0.8$ dex. This is also similar to depletions 
observed either in ``cold'' or ``warm-halo'' clouds \citep{savage96}. 
Note that in this case the warm and halo depletion patterns
can not be differentiated from each other (there is only an estimate of the depletion 
of P in halo gas). Only two values are available within the LB: a solar
value and $-0.3$ dex below solar. 

Plotting the results for [\ion{P}{2}/\ion{Si}{2}] and
[\ion{Fe}{2}/\ion{Si}{2}] for common sightlines
could {\em a priori} differentiate between
the different depletion patterns since the cold, warm, and halo 
depletion patterns have different locii. 
Figure~\ref{depplot} shows  [\ion{P}{2}/\ion{Si}{2}] versus [\ion{Fe}{2}/\ion{Si}{2}]
along with the expected depletion patterns observed in more distant 
sightlines in the Galaxy. Most of the clouds appear to be dominated 
by more warm-like or halo-like gas than cold gas. The scatter 
is probably mainly explained by the fact that the clouds are
not resolved and therefore are a combination of different 
gas phases. The sightline toward WD\,0131$-$163 is more complicated to 
understand because [\ion{P}{2}/\ion{Si}{2}] points toward 
a cold depletion pattern while [\ion{Fe}{2}/\ion{Si}{2}] toward
an halo depletion pattern.

The solar ratios of Si/C, P/C, Fe/C are
$-1.00,-2.98,-1.04 $, respectively. Toward the few sightlines 
where a measurement of \ion{C}{2} was possible, the ratios
([\ion{Si}{2}/\ion{C}{2}], [\ion{P}{2}/\ion{C}{2}], [\ion{Fe}{2}/\ion{C}{2}])
are: $(0.18,<1.40, <-0.13)$ toward WD\,0549+158, 
$(<0.05,<1.64,<-0.06)$ toward WD\,1254+224, $(-0.98,<0.12,-1.62)$ toward WD\,1314+293,
$(-1.14,-0.84,-1.78)$ toward WD\,1634$-$573. The two first sightlines 
do not have any stringent limits. Toward WD\,1314+293, these ratios
are compatible with a cloud having a cold depletion pattern (in agreement 
with the ratios of P/Si and Fe/Si). These ratios for WD\,1634$-$573 
point also to a cold depletion pattern ([\ion{P}{2}/\ion{C}{2}] is, however, 
far too small), but this is in contradiction with the results from 
P/Si and Fe/Si (see Figure~\ref{depplot}), which might suggest that there is
a lack of dust containing C toward this sightline. 

The main reason for the deficiencies of Si and Fe are understood to be due to 
the incorporation of some of the Fe and Si into dust grains. 
The scatter in the values may reflect changes in the dust fraction or 
complications in the ionization fractions that are similar to those
that we found for N, Ar, and O. The fact that we do not observe
any trend with the distance or total column density of the sightline is 
probably mainly due to the fact that the clouds are not resolved. 

\section{Electron Density and Cooling Rate in the LISM}\label{cii}

\subsection{Electron Density}
If collisions with electrons are the principal source of fine-structure excitation of \ion{C}{2} in the LISM, then 
the familiar collisional equilibrium equation between \ion{C}{2} and \ion{C}{2*} \citep[e.g.,][]{gry01}
can be simplified to
\begin{equation}\label{ne_eq}
n_e = 0.189 \, T^{0.5} \frac{N({\rm C}^{+*})}{N({\rm C}^{+})}\;\;\; {\rm cm}^{-3}\,,
\end{equation}
 when $N({\rm C}^{+*}) \ll N({\rm C}^{+})$ and $T \gg 93$ K.
This relation provides an estimate of the electron density when the space 
densities  are replaced by column densities.  
The collisions with neutral hydrogen atoms should
not be significant in the LISM. Indeed, according to Fig.~3 of \citet{keenan86},
$n($\ion{H}{1}$) = 1.0$ cm$^{-3}$ alters the inferred electron density only when
$\log[N({\rm C}^{+*})/N({\rm C}^{+})] \la -2.7$, but our calculations indicate that
$n($\ion{H}{1}$) <0.3 $ cm$^{-3}$ in the LISM as long as $p/k \le 2500$ cm$^{-3}$\,K 
(see Figure~\ref{figcal}). 

There is no evidence of C being depleted into dust in the LISM (see \S~\ref{dust}). 
P is also known to be only lightly depleted. Moreover, in other regions 
of the Galaxy, C and P have similar depletions \citep{savage96}. 
Differences in ionization of C and P are probably not important ($N($\ion{C}{3})
is small with respect to $N($\ion{C}{2}); see \S~\ref{ions} and Figure~\ref{figcal}), 
so that we can write $\log[N($\ion{C}{2*}$)/N($\ion{C}{2}$)] \approx 
\log N($\ion{C}{2*}$) -\log N($\ion{P}{2}$) + \log({\rm P}/{\rm C})_\odot$, 
where the solar ratio $\log({\rm P}/{\rm C})_\odot = -3.0$ dex.

Under these assumptions and assuming a temperature of $T = 7000$ K observed in LIC, 
we can determine the electron density. This is shown in the bottom panel of Figure~\ref{ciiex}. 
The mean value is $n_e \la 0.1$ cm$^{-3}$. Toward Capella situated at a distance
of $d = 12.9$ pc, \citet{wood97} 
determined $n_e = 0.05$--0.23 cm$^{-3}$.
We also plot in  Figure~\ref{ciiex} the expected electron density from our model 
versus the shielding of \ion{H}{1} (see \S~\ref{ions}). 
All the points fall above this line, suggesting that they are probably multiple regions where the
shielding in each region is much less than the total column density of the sightline. 

\subsection{Cooling Rate}
Because carbon has a high abundance, 
one of the most efficient way for cooling the diffuse 
gas is via the excitation of the \ion{C}{2} by collisions
with electrons and hydrogen atoms followed by the spontaneous
emission at 157.7 $\micron$ when the levels decay to the ground
state.
It follows that the \ion{C}{2} cooling rate  
in erg\,s$^{-1}$\, per \ion{H}{1} atom
can be written \citep[e.g.,][]{savage96a},
\begin{equation}\label{cool}
l_c = 2.89 \times 10^{-20}\frac{N({\rm C}^{+*})}{N({\rm H}^{0})}\,,
\end{equation}
where $\log N($\ion{H}{1}$) = \log N($\ion{O}{1}$) + 3.5$ (see \S~\ref{prop}). 
In those conditions, the mean value in our sample is $\log l_c = -25.59 \pm 0.32 $ dex. 
Using \ion{P}{2} as a proxy for the total H (\ion{H}{1}+\ion{H}{2}) and assuming 
no depletion, we 
have $\log l_c = -25.30 \pm 0.39 $ dex in our sample. If a depletion of $\sim -0.2$ to $-0.3$ dex 
is assumed for P \citep{savage96,welty99}, both determinations are in good 
agreement. This result is in agreement with previous determinations derived
in the diffuse gas in other parts of the Galaxy via different methods 
\citep{gry92,bock93,savage96a,boulanger96}. Toward Capella, using 
the \ion{H}{1} and  \ion{C}{2*} column densities  from \citet{wood02b}, 
$\log l_c = -25.14 \pm 0.03 $ dex, but increases by about 0.3 dex 
when using \ion{O}{1} as a proxy of \ion{H}{1}. 

In the top panel of Figure~\ref{ciiex}, we show the cooling rate against
the total \ion{H}{1} column density, showing a clear anti-correlation between
these two quantities. This likely indicates that
for low column density sightlines, a substantial amount of \ion{C}{2} is associated with
ionized gas, which is not taken into account in the normalization by the $N($H) column 
density.

\section{High Velocity Components}\label{hvc}
The sightlines toward WD\,0455$-$282, WD\,1202$+$608, and WD\,1528$+$487 each show 
a high velocity component separated by $-56$, $-40$, and $-39$ 
\km, respectively, with respect to the lower velocity component. In the different
tables, the high velocity component is designated by the letter ``a" after the star name.
Toward WD\,0455$-$282 and WD\,1528$+$487, high velocity components are clearly
detected only in the \ion{C}{2} absorption line for the elements 
considered here. For WD\,0455$-$282, there is also a clear detection of this component
in the \ion{H}{1} Lyman lines. For WD\,1528$+$487, 
there is only a hint in the Lyman lines. This rules out 
misidentification due to a stellar line. For these two stars, 
the presence of  \ion{C}{2} absorption features and the non-detection of neutral 
species, in particular the strong \ion{O}{1} $\lambda$1039
line, implies that these clouds are mostly ionized.  

Toward WD\,1202$+$608, several species are detected (see 
Table~\ref{t4}), making the identification as a high-velocity
component unquestionable. In contrast to the two other
sightlines which have a very low column density with respect
to their low velocity counterparts, this cloud has comparable column
densities compared to the lower velocity components 
for \ion{Si}{2} and \ion{Fe}{2}. However, it has much smaller column densities
for the neutral species, \ion{O}{1} and 
\ion{N}{1}, implying that this cloud is predominantly ionized. In addition,
\ion{N}{1}/(\ion{N}{2}+\ion{N}{1}$)<0.09$, showing that at least
91\% of N is in an ionized form. A similar comment applies for \ion{Ar}{1},
even though the result remains more uncertain for this atom.
Unfortunately, we have no information for \ion{C}{3} and 
\ion{Fe}{3}, because the former is blended with the local 
component (but the large lower limit -- the largest
in our sample -- suggests a fair amount of \ion{C}{3}), and 
\ion{Fe}{3} is confused by the moving stellar lines from this binary system.
The column density of \ion{O}{1} suggests  
an \ion{H}{1} column density of $\sim 18.20$ dex, while \ion{Si}{2}
gives a total H column density of $>18.6$ dex (it is a lower
limit because Si can be substantially depleted onto dust grains), implying an
\ion{H}{2} column density $>18.3$ dex.

\citet{holberg95} and \citet{kruk02a} argue that the origin of the high velocity
component toward WD\,1202$+$608 may be a remnant from the 
ejected common envelope of the binary, assuming it is a relatively young
system. \citet{gry01} also identified two negative high velocity components
in the line of sight toward $\epsilon$\,CMa at $-27$ \km\ and $-82$ \km\ from the
main component. The fact that the velocity of these components is
always negative would also point toward a circumstellar origin. 
However, it should be noted that the high velocity features appear in pairs of 
stars.  WD\,0455$-$282 is in the same part of the sky as $\epsilon$ CMa, and 
WD\,1202$+$608 is in the same part of the sky as WD\,1528$+$487.  This suggests 
that in each of these two cases, the sightlines could be passing through 
a single gas cloud. If so, the clouds are probably much closer than the 
stars and because of the high ionization are likely well within the LB.
Although the argument is not definitive, it does weigh 
against the hypothesis of stellar ejecta.  Finally, we note that this study does 
not show widespread high-velocity components along many sight lines.  In 
particular, there is no evidence of globally expanding gas at positive 
velocities as might be expected for the case of a recent supernova explosion positioned near the Sun.

\section{Summary and Concluding Remarks}
We present a comprehensive survey of the LISM which includes 
the main species available in the far-uv wavelength 
range (\ion{C}{2}, \ion{C}{2*}, \ion{C}{3}, 
\ion{N}{1}, \ion{N}{2}, \ion{O}{1}, \ion{Ar}{1}, \ion{Si}{2}, 
\ion{P}{2}, \ion{Fe}{2}, \ion{Fe}{3}, and H$_2$). 
The spectral resolution ($R\sim 20,000$) is not sufficient for
detailed analysis of velocity structures. 
However, the number of species available permit a study of some of the  
fundamental characteristics of the local interstellar 
gas by comparing the relative column densities,
the ionization structure, the molecular 
content, and the indirect indication of dust.

\begin{figure}[tbp]
\epsscale{1}
\plotone{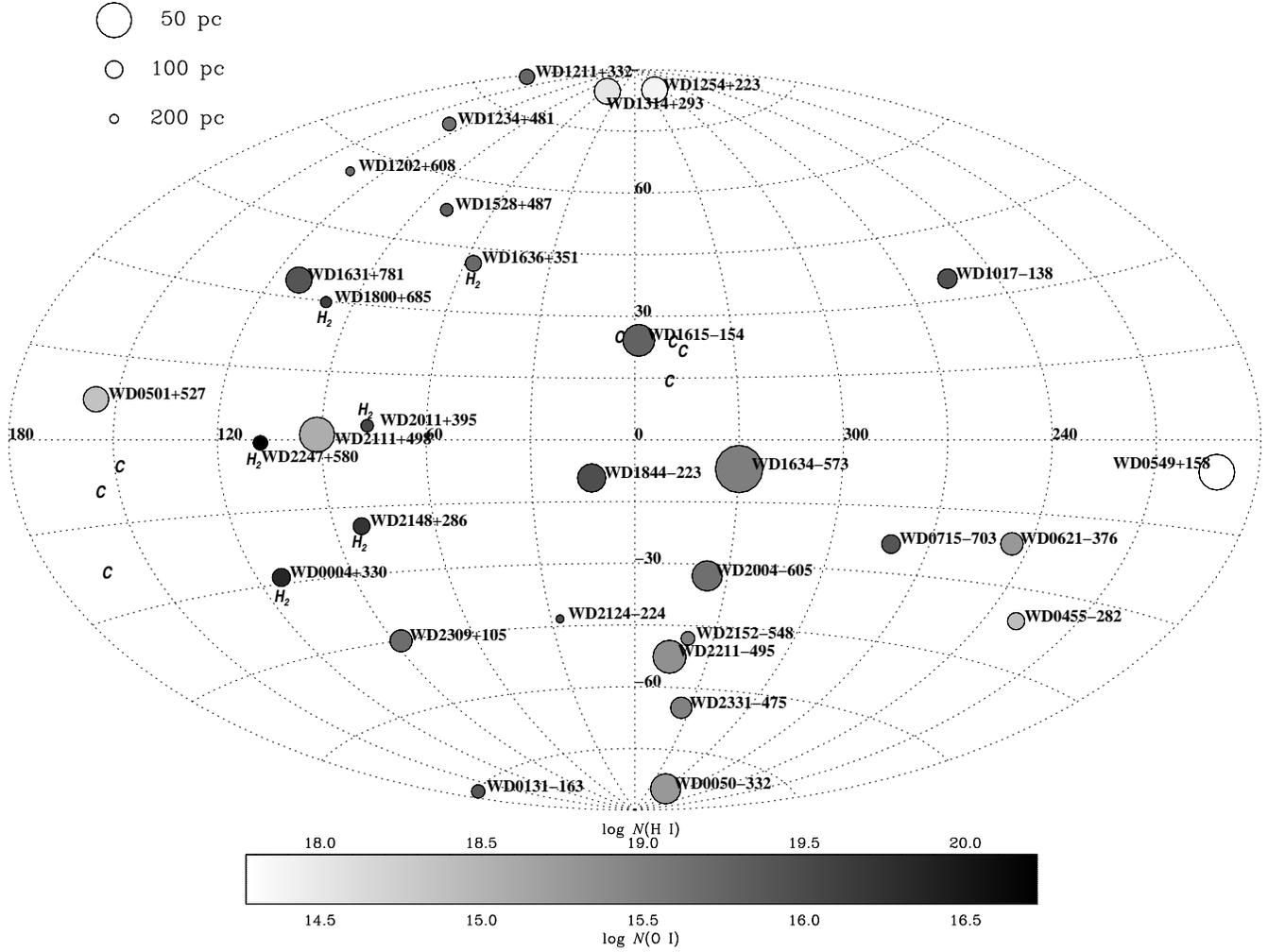}
\caption{Target locations  and the total \ion{O}{1} column densities
in  Galactic coordinates ($l,b$). 
The diameter of each circle is inversely proportional to  the distance of the line of sight, 
and the shading of the symbol indicates the total column density. 
The \ion{H}{1} values were computed from the \ion{O}{1} values using the \ion{O}{1}/\ion{H}{1} 
ratio discussed in the text.  We indicate the 
presence of molecular hydrogen by the symbol H$_2$ for detection by this survey and ``C'' for sightlines
near WD\,1615$-$154 and $l\sim 150\degr$ from the survey 
of  \citet{spitzer74} based on {\em Copernicus} data (the stars  in the {\em Copernicus} study 
are between 111 and 166 pc). 
\label{oimap}}
\end{figure}

\begin{figure}[tbp]
\epsscale{1}
\plotone{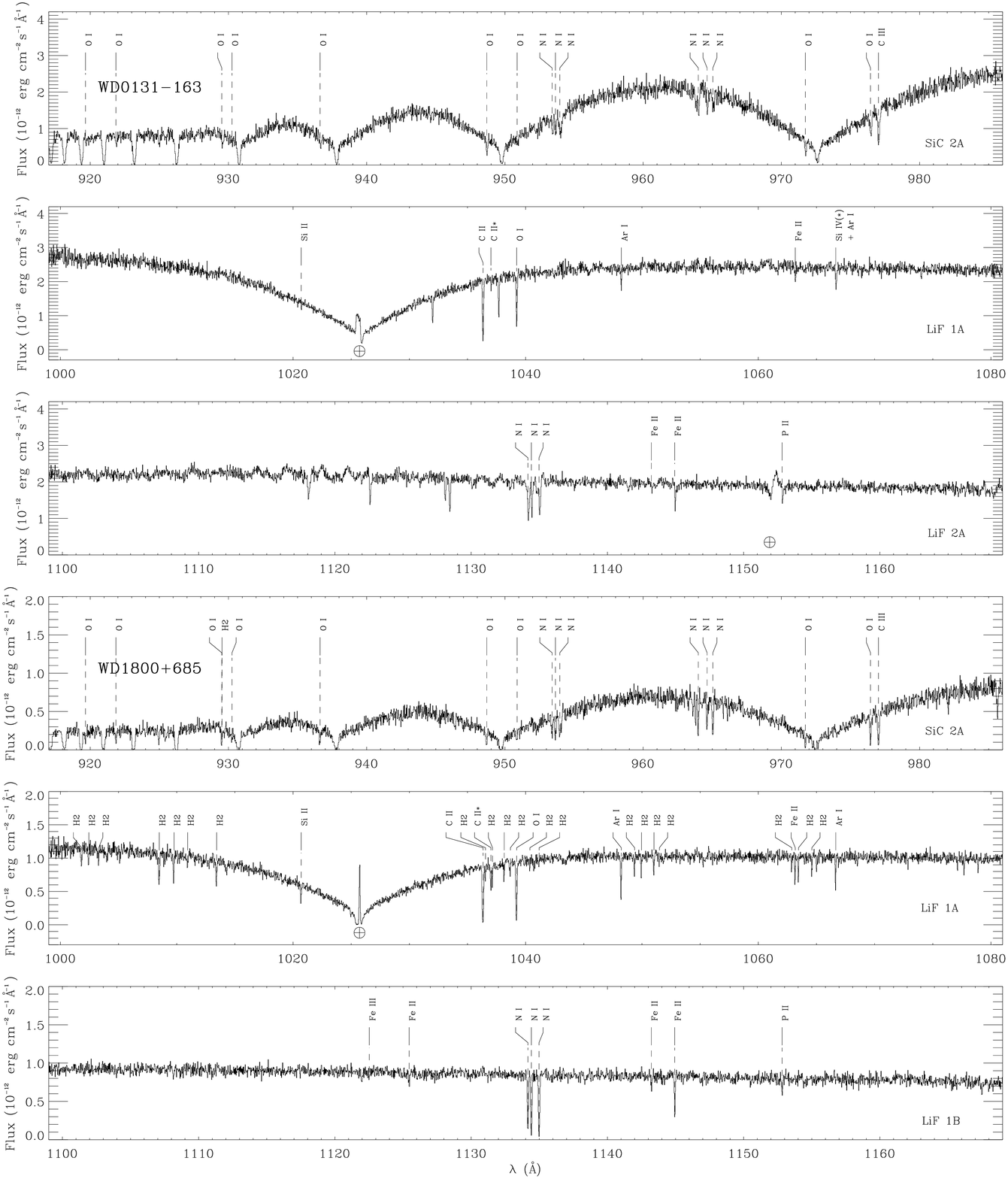}
\caption{Two examples of {\fuse} WD spectra, with WD\,0131$-$163 shown in
  top three panels and  WD\,1800$+$685 in bottom three panels. 
   WD\,1800$+$685 is a pure hydrogen WD with no photospheric metal
   absorption lines, but shows absorption by interstellar atoms, ions, and
   H$_2$. WD\,0131$-$163 
  has a few stellar lines (note in particular the stellar \ion{Si}{4}
  blended with \ion{Ar}{1} at 1066.66 \AA). 
  The positions of strong airglow emission lines  
  are indicated by the earth symbol.
\label{spec}}
\end{figure}

\begin{figure}[tbp]
\epsscale{0.7}
\plotone{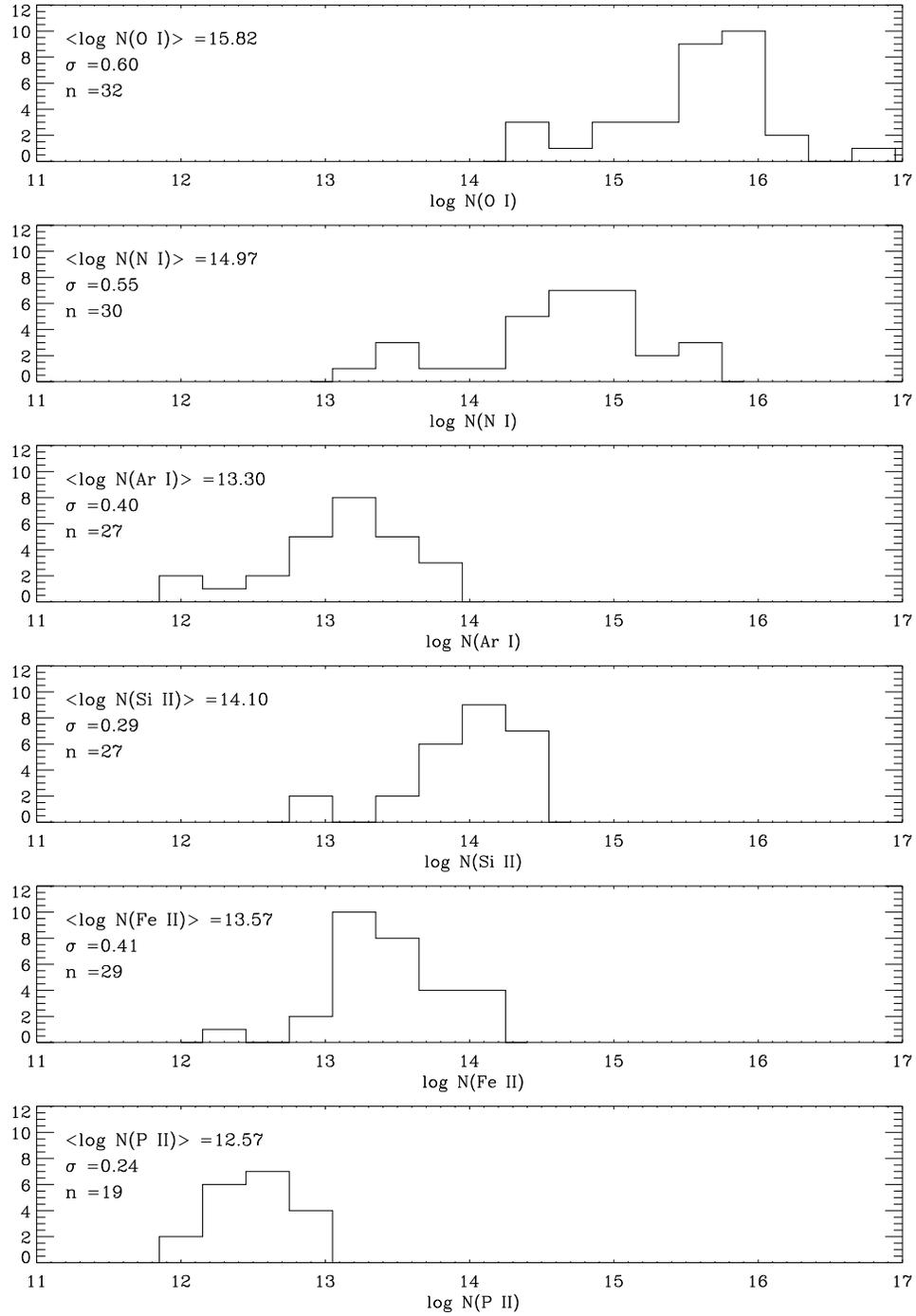}
\caption{Histograms of column densities for the different ions observed. The mean, standard deviation and
number of detected clouds ($n$)  are listed in the upper-left of each plot. 
\label{comphist}}
\end{figure}

\begin{figure}[tbp]
\epsscale{1}
\plotone{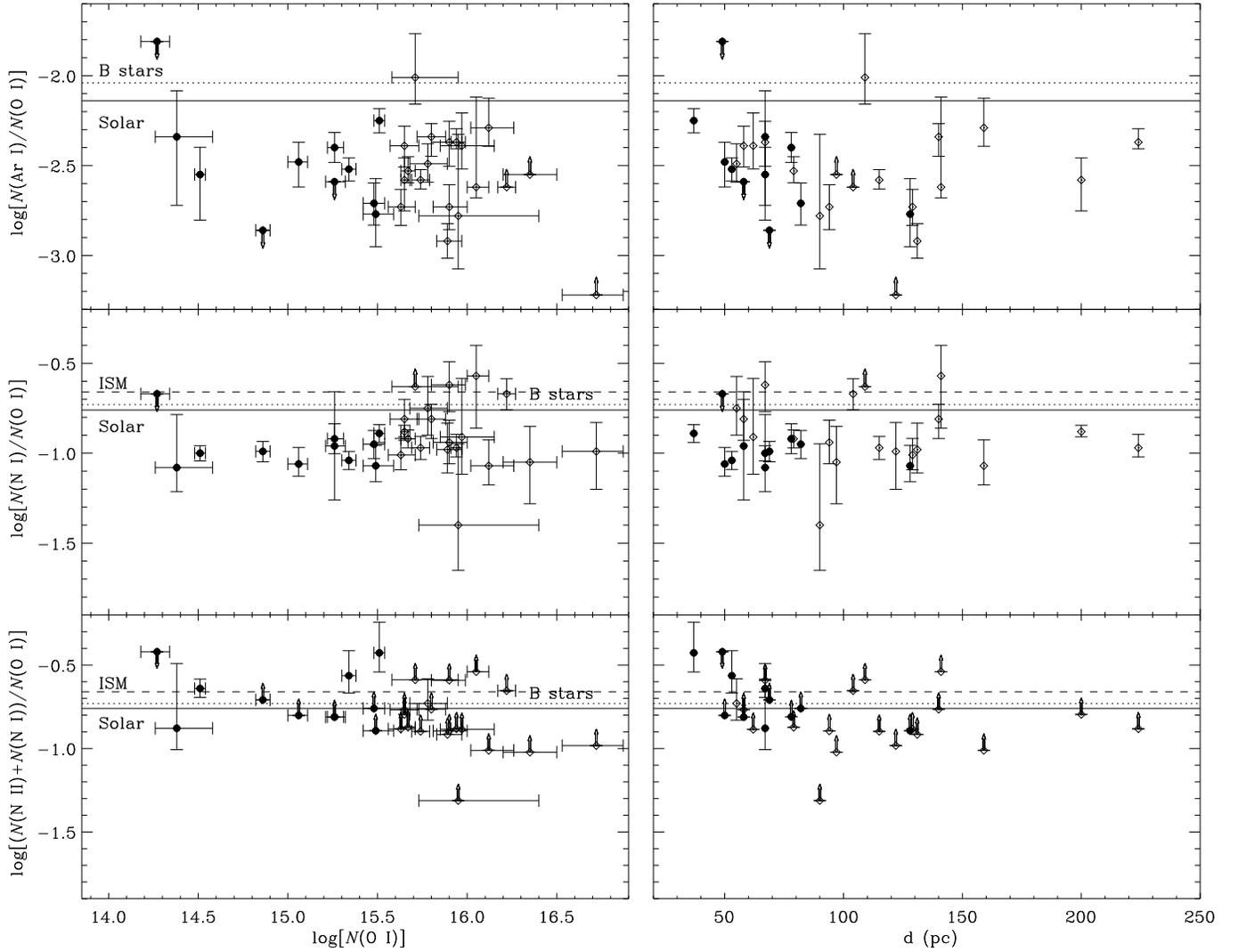}
\caption{Logarithmic ratios of 
  \ion{Ar}{1}, \ion{N}{1}, and (\ion{N}{1}\,+\,\ion{N}{2})/\ion{O}{1} plotted versus the total \ion{O}{1}
  column density and the distance of the sightlines
  (note that the distance is only an upper limit to the interstellar clouds). The horizontal lines
  indicate the solar, B-type star, and interstellar values.  The Local Bubble ($\log N($\ion{O}{1}$)\la 15.60$) data points 
  are shown with filled circles.
\label{comp}}
\end{figure}

\begin{figure}[tbp]
\epsscale{0.7}
\plotone{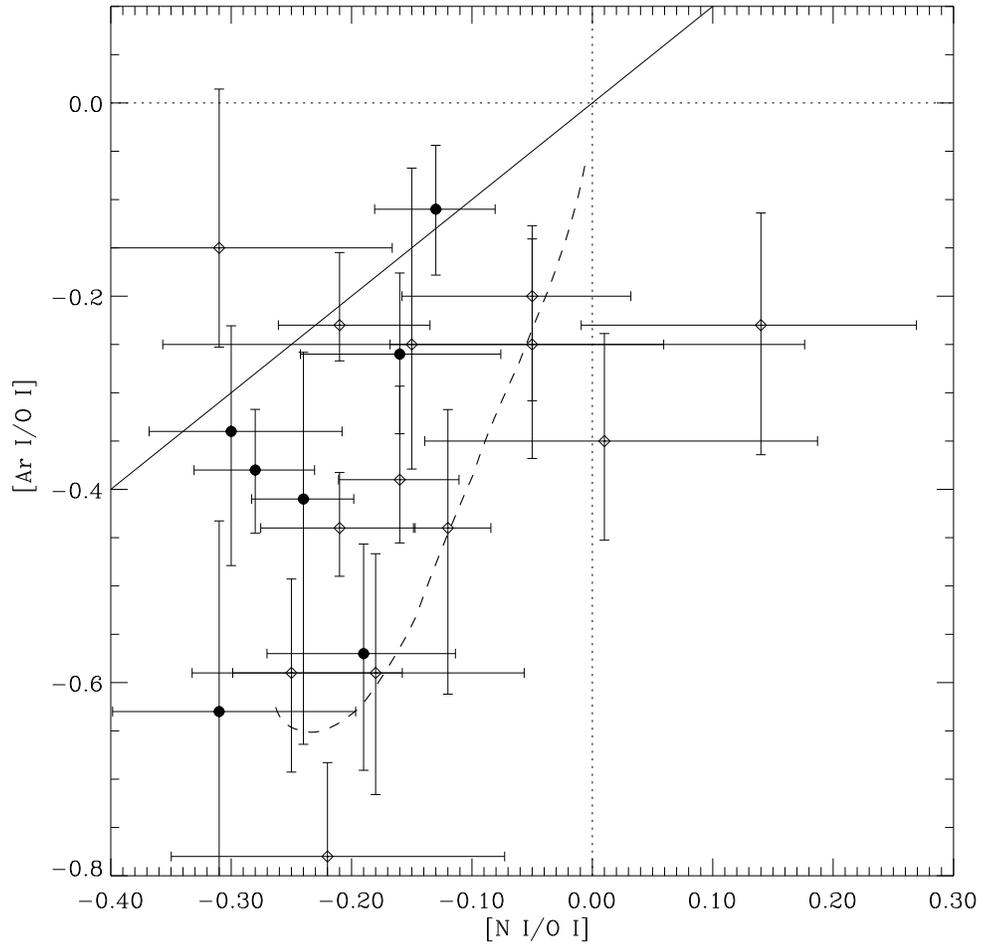}
\caption{Logarithmic ratio of \ion{Ar}{1}/\ion{O}{1} plotted versus \ion{N}{1}/\ion{O}{1}. 
Both ratios are normalized
to their solar values. The straight continuous line shows a 1:1 relationship. The dashed 
lines present the expected deficiency of \ion{Ar}{1} and \ion{N}{1} 
from a photoionization model (see \S~\ref{model} and Figure~\ref{figcal}). 
The Local Bubble ($\log N($\ion{O}{1}$)\la 15.60$) data points are shown with filled circles.
\label{ratio}}
\end{figure}

\begin{figure}[tbp]
\epsscale{0.7}
\plotone{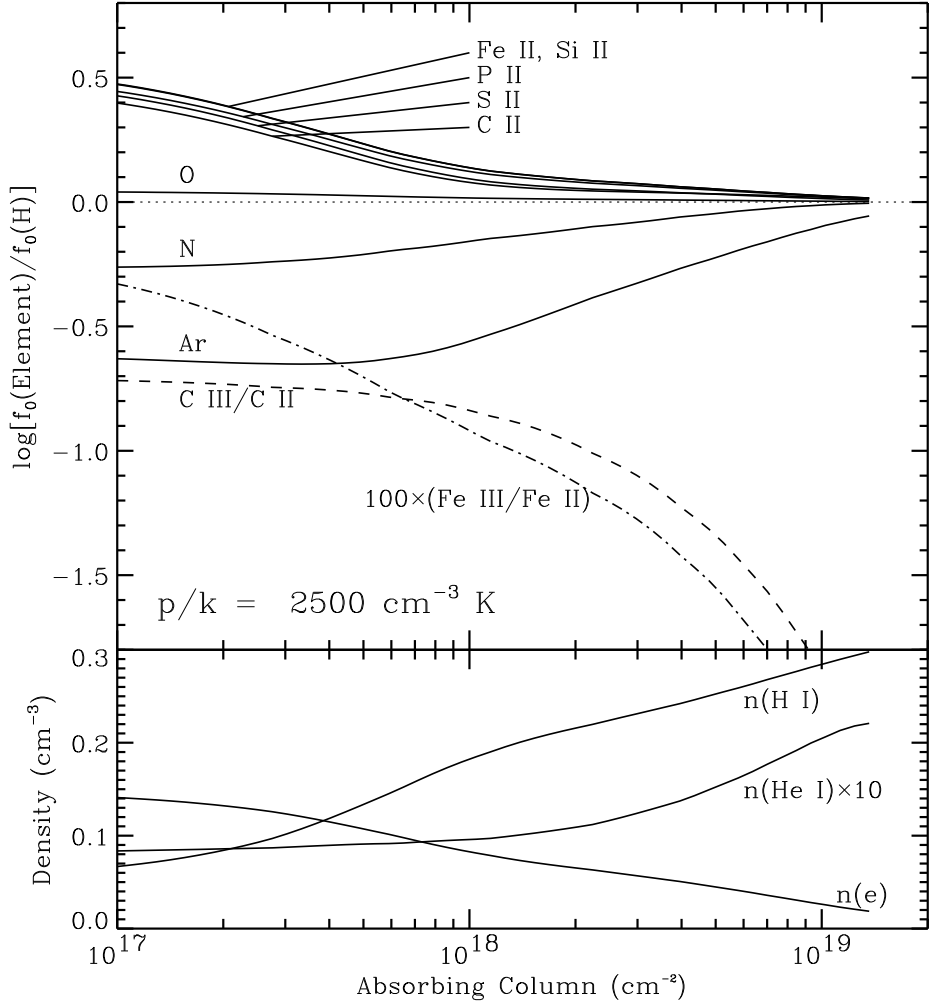}
\caption{Results from a photoionization model of gas in the LB. 
{\em Top}: Logarithms of the expected deficiencies in the 
neutral fractions of $f_0 = n({\rm neutral})/n({\rm total})$ of N, O,
and Ar compared to those of H (see \ref{model}
for more details). Also plotted are the logarithms of
 $n({\rm X^+})/n({\rm total})$, i.e  the singly
ionized species, compared to the neutral fraction of H, along with the ratios \ion{C}{3}/\ion{C}{2} 
and \ion{Fe}{3}/\ion{Fe}{2} (using the same logarithmic scale on the 
vertical axes). All the calculations assume the same ionizing 
radiation field and include the effects of charge exchange with neutral hydrogen
These fractions are plotted as 
a function of absorbing column and hence shielding depth scaled to $N($\ion{H}{1}) 
for a cloud of uniform pressure $p/k$.  {\em Bottom}:
Densities of neutral hydrogen, neutral helium, and electrons against
shielding depth.
\label{figcal}}
\end{figure}

\begin{figure}[tbp]
\epsscale{1}
\plotone{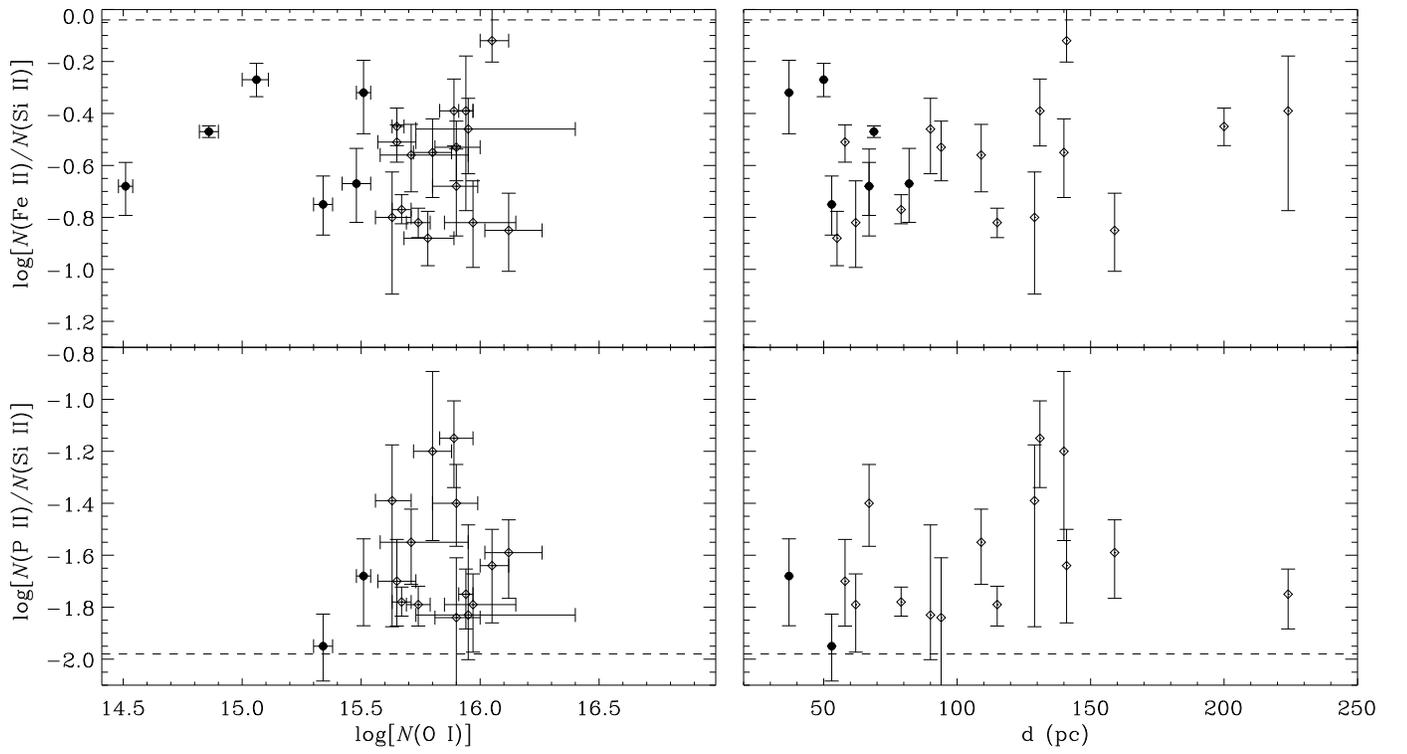}
\caption{Logarithmic ratios of 
of \ion{Fe}{2} and \ion{P}{2} to \ion{Si}{2} against the total \ion{O}{1}
column density and against the distance of the sightlines.
The dashed lines indicate the solar value of those ratios. 
The Local Bubble ($\log N($\ion{O}{1}$)\la 15.60$) data points 
  are shown with filled circles.
\label{compsifep}}
\end{figure}

\begin{figure}[tbp]
\epsscale{0.7}
\plotone{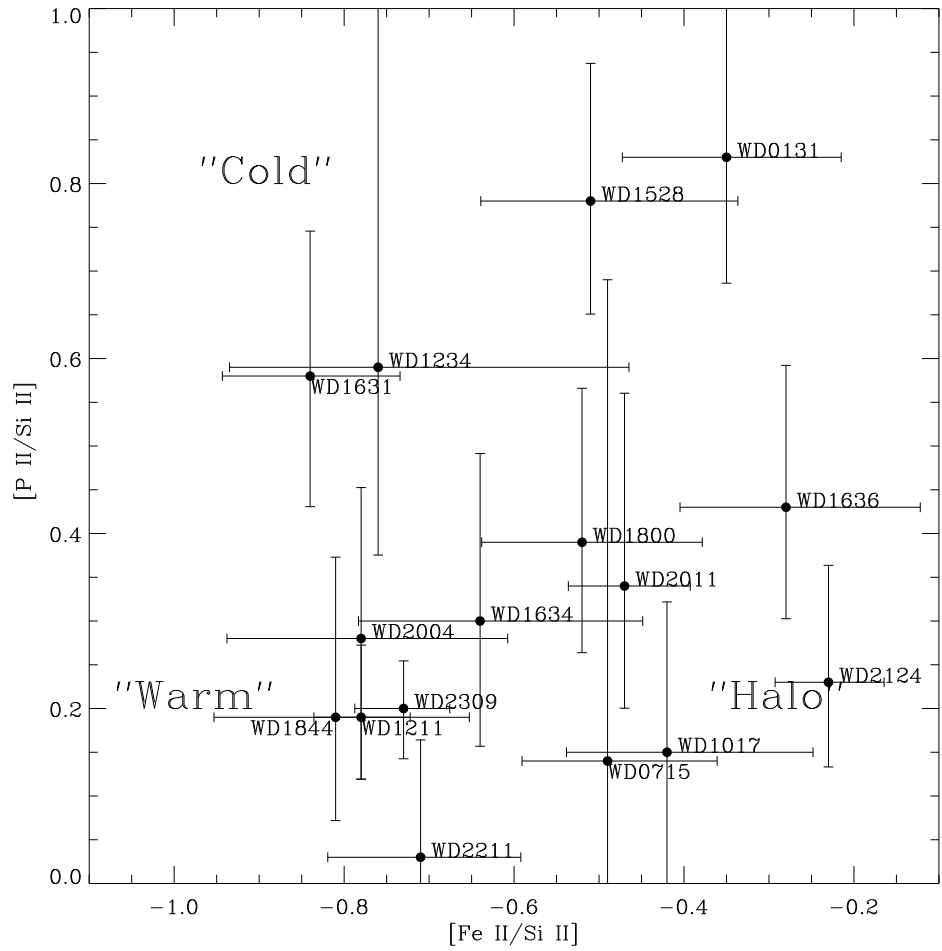}
\caption{Logarithmic ratio of \ion{P}{2}/\ion{Si}{2} against \ion{Fe}{2}/\ion{P}{2} normalized
to their solar values. The ``Cold'', ``Warm'', and ``Halo'' refer to depletion patterns
observed in our Galaxy. We also indicate for each point the first 4 digits of the WD. 
\label{depplot}}
\end{figure}

\begin{figure}[tbp]
\epsscale{0.7}
\plotone{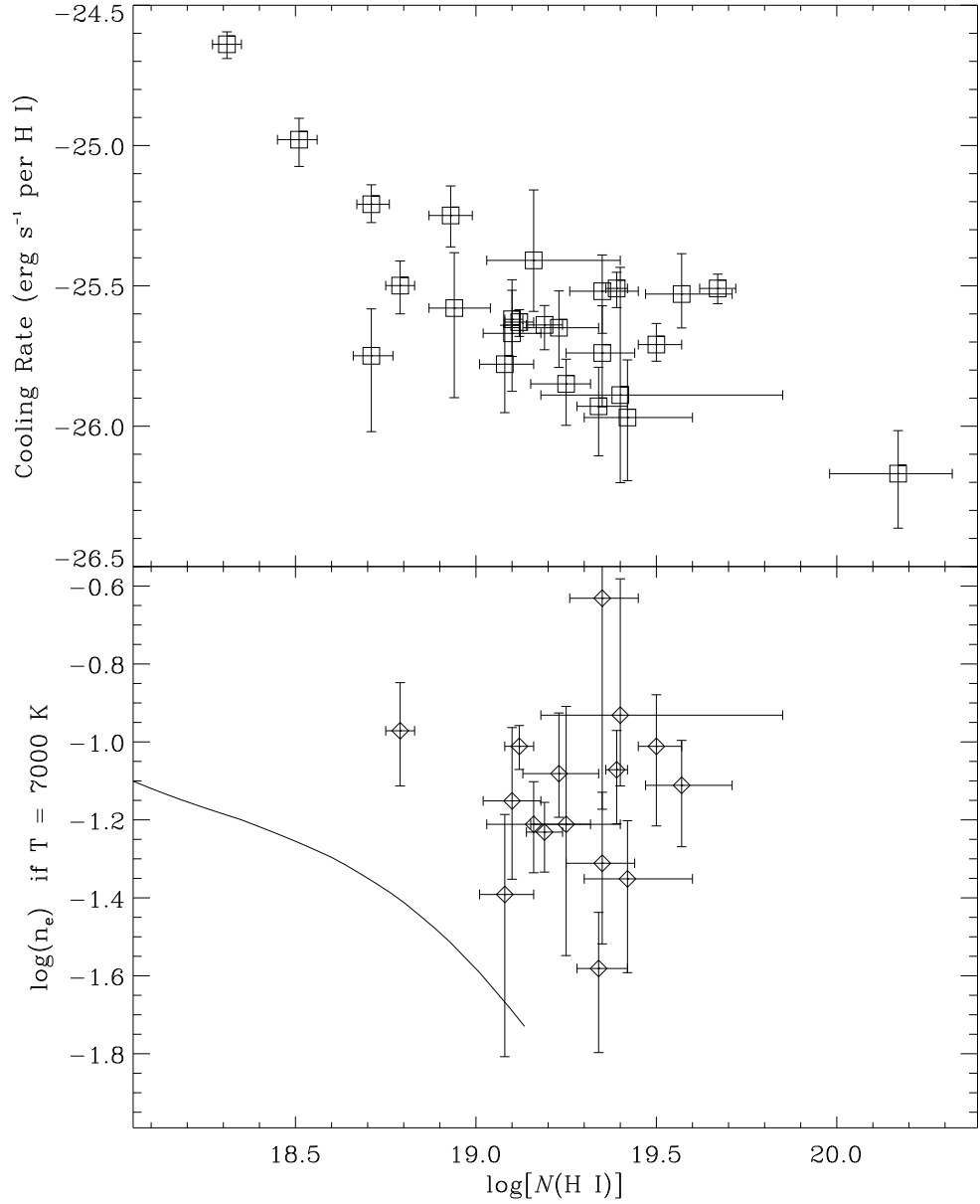}
\caption{Logarithmic of the cooling rate in erg\,s$^{-1}$\, per \ion{H}{1} atom (upper panel) and of the 
electron density in cm$^{-3}$ (lower panel) against the total \ion{H}{1}
column density (the latter being derived from \ion{O}{1}: $\log N($\ion{H}{1}$) = \log N($\ion{O}{1}$) + 3.5$)
(see \S~\ref{cii} for more details). For clarity, we do not show the lower limits, but none contradicts
the measured values. Also shown as a solid line is $\log n_e$ from Figure~\ref{figcal}.
\label{ciiex}}
\end{figure}

Our survey detected only very 
diffuse molecular clouds ($f_{\rm H_2} \la 10^{-4}$). 
H$_2$ was detected in only six out of 31 sightlines all within a single
sector of the sky (see Figure~\ref{oimap}).
Thus it appears that H$_2$ is not widely distributed in the LISM. 
The derived H$_2$ $b$-values for the 6 sightlines in our sample range from 2.8 to 5 \km,
implying  thermal temperatures less than 900--3000 K. 
This is different from the temperature of 7000 K usually observed in 
local interstellar clouds, implying different gas phases in the LISM. 
Combining our results with previous work on H$_2$ in the LISM shows
that the molecular phases are not uniform. Molecular 
hydrogen can exist in diffuse clouds, but in some regions the molecular fractions
are similar to the atomic content. 

The comparison of column densities of \ion{Si}{2}, \ion{P}{2}, 
\ion{Fe}{2} provides information on the dust depletions 
of LISM. The relative abundances of these species 
vary from sightline to sightline, again suggesting  
different gas-phases in the LISM. These abundance ratios
are better represented by typical warm or halo diffuse
clouds observed in more distant sightlines in the Galaxy rather
than a cold depletion pattern. 

The cooling rate in the LISM, $l_c$ (in erg\,s$^{-1}$\, \ion{H}{1} atom$^{-1}$, 
where the \ion{H}{1} column density was determined from the \ion{O}{1} column
density), can be derived directly from the \ion{C}{2*} absorption lines and 
has a mean value of $\log l_c = -25.59 \pm 0.32 $ dex, very similar to 
previous determinations. Higher values are observed at lower \ion{H}{1} column densities, 
indicating that for low column density sightlines, a substantial amount of the \ion{C}{2} 
must come from ionized gas. The electron density can also be inferred from 
the \ion{C}{2*} absorption lines. Using \ion{P}{2} as a proxy of 
\ion{C}{2} and assuming a temperature of 7000 K, the electron density
in the LISM is $n_e \la 0.1$ cm$^{-3}$, but larger than the
values predicted by our model. This disagreement likely indicates that
there are multiple regions along the sightlines where the \ion{H}{1} shielding in each region
is much less than the total \ion{H}{1} column density of the sightline.

The comparison of column densities of \ion{N}{1} and \ion{Ar}{1} to
\ion{O}{1} provides information on the ionization structure 
of the LISM. The \ion{N}{1}/\ion{O}{1} ratio is systematically $-0.24$
dex subsolar in the LB. Near or beyond the LB boundary, 
the picture is far more complicated, with  a scatter 
of \ion{N}{1}/\ion{O}{1} between $\sim -0.3$ dex subsolar to 
$\sim +0.15$ dex supersolar (WD\,1636+351). Using measured values and lower
limits on the \ion{N}{2} column densities, we show that
the observed deficiencies are likely due to ionization and not 
abundance variations or dust depletions.
For most sightlines, the \ion{Ar}{1}/\ion{O}{1} ratios also show  deficiencies of 
\ion{Ar}{1} in the LB by $\sim -0.35$ dex with respect to the solar value, 
but with a larger scatter than for \ion{N}{1}, and this scatter increases beyond the LB ranging from 
a solar value to $\sim -0.8 $ dex subsolar.

More highly ionized species such as \ion{C}{3} and \ion{Fe}{3} are also available to constrain the 
ionization structure. \ion{C}{3} appears to be ubiquitous in the 
LISM for our sample, and \ion{C}{3}/\ion{C}{2} appears to vary from sightline to sightline.
The \ion{C}{3}/\ion{C}{2} ratio is however generally small; 
\ion{C}{3} is not a dominant ion in the LISM.
\ion{Fe}{3} remains more elusive and is often completely 
lost in the stellar profile. But our limits on \ion{Fe}{3}/\ion{Fe}{2} 
indicate that \ion{Fe}{2} is the dominant ion in the LISM.

We modified the photoionization code used by \citet{sofia98} and \citet{jenkins00}. 
In particular, we adopted the new radiation field calculated by \citet{slavin98}
and added the atomic parameters for C, P, Si, S, and Fe. 
The calculations were performed for a sequence of models where the thermal
pressure was set to a value of $p/k = 2.5 \times 10^3$ cm$^{-3}$\,K.
These calculations show that \ion{O}{1} follows \ion{H}{1} very well, while
\ion{N}{1} and \ion{Ar}{1} become progressively more deficient as 
the shielding from \ion{H}{1} becomes smaller. 
The other elements (C, P, Si, S, and Fe) remain primarily in the singly 
ionized state. 
The assumed pressure also seems to give a reasonable set of values for  
\ion{C}{3}/\ion{C}{2}. However, the values calculated for \ion{Fe}{3}/\ion{Fe}{2} appear to be
too low, although we have only one measured value of \ion{Fe}{3}/\ion{Fe}{2}, 
the rest are upper limits. 
According to our model, the fraction of \ion{Fe}{3} is heavily dependent 
on the effect of charge exchange with neutral hydrogen and is therefore 
not a reliable tracer of ionized gas. 

\ion{Ar}{1}, \ion{N}{1} and \ion{H}{1} have similar 
recombination coefficients, so if the LISM was highly ionized at an earlier epoch 
by a strong shock and is in the process of recombining, 
these elements would have roughly equal ionized fractions. Our study
contradicts this proposition by indicating that photoionization is a major contribution.
We note, however, that in a few cases, [\ion{N}{1}/\ion{O}{1}] is 
approximatively equal to [\ion{Ar}{1}/\ion{O}{1}], 
suggesting that the ionization fraction of N could be as large as that of Ar, 
which contradicts our model. This could indicate the radiation field might be 
harder in some directions. However, our model does not provide a clear 
explanation for this phenomenon with respect to the different observational 
constraints, and other factors are needed to explain this condition
(e.g., non-equilibrium collisional ionization or overabundance of Ar possibly due to
a supernova).

The variations observed in the ratios of \ion{Ar}{1}/\ion{O}{1}, 
\ion{N}{1}/\ion{O}{1}, \ion{N}{1}/(\ion{N}{1}+\ion{N}{2}), and \ion{C}{3}/\ion{C}{2}
indicate that the photoionization conditions are probably not uniform in the LB and 
in the LISM. Such changes in the photoionization conditions might be linked to 
the patchiness observed in the distribution \ion{O}{6} absorption \citep{oegerle03}.
These variations, in turn, may depend on the strength and direction of the 
magnetic field at the cloud boundaries. 
\citet{slavin02} argued that evaporative interfaces at the boundaries of the clouds
(\ion{O}{6} absorption being a tracer of such interfaces) could account for a large
fraction of the observed ionization. Better observational constraints will come from 
higher spectral resolution of far-UV spectra, in which the individual clouds can 
be resolved. 

Finally, we note that our results do not strongly contradict the X-ray data. 
\citet{juda91} compared the Be and B band fluxes in a number of different
directions and found that the ratio of the two did not change with the
overall intensity.  From this nearly linear relationship between Be and B band
fluxes, they concluded that interspersed neutral gas within the LB must
have an average density less than 0.025 cm$^{-3}$. Unfortunately, their 
sky coverage was limited and most of their viewpoints were not near any of our targets. 
An exception is WD\,0501$+$527, for which $\log N($\ion{H}{1}$)= 18.11$ \citep{lemoine02}, 
and $d = 69$ pc, which gives $\langle n ($\ion{H}{1}$)\rangle = 0.007$ cm$^{-3}$, below
their limits. But only 30\degr away from one of the X-ray directions, 
toward WD\,1631+781, the same calculation gives 0.1 cm$^{-3}$. It would
be interesting to see if toward denser lines of sight (e.g., WD\,0004+330 
gives  $\langle n ($\ion{H}{1}$)\rangle = 0.24$ cm$^{-3}$), there is 
measurable X-ray absorption in the lowest energy band.

\acknowledgments
We thank Cristina Oliveira, 
Guillaume H\'ebrard, Jean Dupuis, Martial Andr\'e, Jeff Kruk, and Bill Oegerle
for allowing us to use their results prior to publication and for enlightening discussions. 
We thank Jeff Linsky, Seth Redfield, and Barry Welsh for a careful reading of an 
earlier version and useful suggestions. 
This work is based on data obtained
for the Guaranteed Time Team by the NASA-CNES-CSA FUSE mission operated 
by the Johns Hopkins University. Financial support to U. S.
participants has been provided by NASA contract NAS5-32985.
French participants are supported by CNES.
This research has made use of the NASA
Astrophysics Data System Abstract Service and the SIMBAD database,
operated at CDS, Strasbourg, France.

\begin{deluxetable}{llccccllc}
\tablecolumns{9}
\tablewidth{0pc} 
\tabletypesize{\scriptsize}
\tablecaption{Sightline summary \label{t1}} 
\tablehead{\colhead{WD name}    & \colhead{Other name}    &   \colhead{$l$}&   \colhead{$b$}&   \colhead{$v_{\rm helio}$}&   \colhead{$d$}&\colhead{Spectral}& \colhead{Aperture} &\colhead{Notes}  \\
\colhead{}    & \colhead{}    &   \colhead{(\degr)}&   \colhead{(\degr)}&   \colhead{(\km)}  &  \colhead{(pc)}&  \colhead{Type}&  \colhead{}&  \colhead{}}
\startdata
WD\,0004$+$330 & GD 2         	&	 112.48  &    $ -28.69 $ &  $ +0.1 $   		&     97$^a$ 	 	    &  DA     &  LWRS, MDRS	     & [A]  (1)    \\
WD\,0050$-$332 &GD 659        	&	 299.15  &    $ -84.12 $ &  $ +12.2 $  		&     58$^a$	 	    &  DA1    &  LWRS		     & [B]  (2)    \\
WD\,0131$-$163 &GD 984        	&	 167.26  &    $ -75.15 $ &   \nodata   		&     96$^a$	 	    &  DA1    &  LWRS		     & [B]  (2)    \\
WD\,0455$-$282 &MCT 0455-2812 	&	 229.29  &    $ -36.17 $ &  $ +13.6 $  		&     102$^a$	 	    &  DA1    &  MDRS		     & [C]  (2)    \\
WD\,0501$+$527 &G191-B2B      	&	 155.95  &    $ +7.10  $ &  $ +13.0 $  		&     $69 \pm 16$ (59$^a$)  &  DA1    &  LWRS, MDRS, HIRS    & [C]  (3)    \\
WD\,0549$+$158 &GD 71         	&	 192.03  &    $ -5.34  $ &  $ +23.2 $  		&     49$^a$	 	    &  DA1    &  LWRS		     & [A]  (2)    \\
WD\,0621$-$376 &   		&	 245.41  &    $ -21.43 $ &  $ +19.5 $  		&     78$^a$	 	    &  DA     &  LWRS, MDRS	     & [D]  (4)    \\
WD\,0715$-$703 &   		&	 281.62  &    $ -23.50 $ &   \nodata   		&     94$^a$	 	    &  DA1    &  MDRS		     & [A]  (2)    \\
WD\,1017$-$138 &   		&	 255.74  &    $ +34.53 $ &  $ +34.5 $  		&     90$^a$	 	    &  DA     &  LWRS		     & [A]  (2)    \\
WD\,1202$+$608 &GD 314, Feige 55&	 133.11  &    $ +55.66 $ &  $ -16.1 $  		&     200$^d$	 	    &  DA     &  MDRS		     & [C]  (2)    \\
WD\,1211$+$332 &HZ 21 		&	 175.03  &    $ +80.02 $ &  $ -18.0 $  		&     $115 \pm35 $ 	    &  DO     &  LWRS, MDRS	     & [B]  (5)    \\
WD\,1234$+$481 &   		&	 129.81  &    $ +69.01 $ &  $ -28.9 $  		&     129$^a$	 	    &  DA     &  LWRS		     & [A]  (2)    \\
WD\,1254$+$223 &GD 153    	&	 317.26  &    $ +84.75 $ &  $ -11.1 $  		&     67$^a$	 	    &  DA1    &  LWRS		     & [A]  (2)    \\
WD\,1314$+$293 &HZ 43    	&	 54.10   &    $ +84.16 $ &  $ -17.4 $  		&     $68 \pm13$ 	    &  DA1    &  LWRS, MDRS	     & [A]  (6)    \\
WD\,1528$+$487 &   		&	 78.87   &    $ +52.72 $ &   \nodata   		&     140$^a$	 	    &  DA:    &  LWRS		     & [B]  (2)    \\
WD\,1615$-$154 &EGGR 118    	&	 358.79  &    $ +24.18 $ &  $ -38.2 $  		&     55$^d$   		    &  DA1.5  &  MDRS		     & [A]  (2)    \\
WD\,1631$+$781 &1ES 1631+78.1   &	 111.29  &    $ +33.58 $ &  $ -11.8 $  		&     67$^a$	 	    &  DA1    &  MDRS		     & [A]  (7)    \\
WD\,1634$-$573 &HD 149499 B     &	 329.88  &    $ -7.02  $ &  $ -19.6 $$^\alpha$  & $37 \pm 3$    	    &  DOZ1   &  LWRS, MDRS	     & [B]  (8)    \\
WD\,1636$+$351 &   		&	 56.98   &    $ +41.40 $ &   \nodata   		&     109$^a$	 	    &  DA     &  LWRS		     & [A]  (2)    \\
WD\,1800$+$685 &   		&	 98.73   &    $ +29.78 $ &  $ -15.9 $  		&     159$^a$	 	    &  DA1    &  MDRS		     & [A]  (2)    \\
WD\,1844$-$223 &   		&	 12.50   &    $ -9.25  $ &   \nodata   		&     62$^a$	 	    &  DA:    &  LWRS		     & [A]  (2)    \\
WD\,2004$-$605 &   		&	 336.58  &    $ -32.86 $ &  $ -28.0 $  		&     58$^a$	 	    &  DA     &  MDRS		     & [A]  (7)    \\
WD\,2011$+$395 &EUVE J2013+40.0 &	77.00	 &    $+3.18   $ &   \nodata   		&     141$^a$	 	    &  DAO    &  LWRS		     & [B]  (2)    \\
WD\,2111$+$498 & GD 394   	&	 91.37   &    $ +1.13  $ &  $ -6.2  $  		&     50$^a$	 	    &  DA2    &  LWRS		     & [B]  (2)    \\
WD\,2124$-$224 &   		&	 26.81   &    $ -43.19 $ &   \nodata   		&     224$^b$	 	    &  DA     &  MDRS		     & [B]  (2)    \\
WD\,2148$+$286$^\ast$ &BD+28\degr4211  &	81.87	 &    $-19.29  $ &   $ -20.2 $$^\beta$ 	& $104 \pm 18$ 		    &  SdO    &  LWRS, MDRS	     & [E]  (9)    \\
WD\,2152$-$548 &   		&	 339.73  &    $ -48.06 $ &   \nodata   		&     128$^a$	 	    &  DA:    &  MDRS		     & [A]  (2)    \\
WD\,2211$-$495 &  	 	&	 345.79  &    $ -52.62 $ &  $ -0.7  $  		&     53$^a$	 	    &  DA     &  LWRS, MDRS, HIRS    & [D]  (10)   \\
WD\,2247$+$583 &Lan 23   	&	107.64   &    $-0.64   $ &   \nodata   		&     122$^a$	  	    &  DA     &  lWRS		     & [C]  (5)    \\
WD\,2309$+$105 &GD 246   	&	87.26	 &    $-45.11  $ &  $-7.9   $  		&     79$^a$	 	    &  DA1    &  LWRS, MDRS	     & [B]  (5)    \\
WD\,2331$-$475 &MCT 2331-4731   &	 334.83  &    $ -64.81 $ &  $ +14.9 $  		&     82$^a$	 	    &  DA     &  lWRS, MDRS	     & [B]   (5)   \\
\enddata
\tablecomments{Heliocentric velocities are from \citet{holberg98}, except 
($\alpha$) from \citet{wood02} and ($\beta$) from \citet{sonneborn02}.
$^\ast$Note that using the denomination  WD\,2148$+$286 for 
BD+28\degr4211 is somewhat incorrect because this star
is a subdwarf of type O, not a white dwarf. However, 
for consistency and simplicity, we have used this name
throughout the article.
$a$: Distances from \citet{vennes97}. 
$b$: Distance from \citet{vennes98}. 
$c$: Distance from Dupuis (2002, private communication).
$d$: Distances from \citet{holberg98}. 
Distances with errors are from Hipparcos parallax measurements \citep{perryman97}; 
for the others, the errors are typically 20--30\%.
Metallicity of the star: [A] no metal detected in {\fuse}\ spectra; [B] few metals; [C] rich in metals; [D] very rich in metals; [E] solar metallicity 
(see Dupuis et al. 2003, in preparation, and references below).
Source of the column densities: (1) Oliveira et al. (2003, private communication), this work; (2) this work; (3) \citet{lemoine02,andre02b};
(4) \citet{lehner02}, this work; (5) \citet{oliveira02}, this work; (6) \citet{kruk02}; 
(7) \citet{hebrard02b}, this work; (8) \citet{wood02}, this work; (9) \citet{sonneborn02}, this work;
(10) \citet{hebrard02a}, this work.
}
\end{deluxetable}

\begin{deluxetable}{lccl}
\tablecolumns{4}
\tablewidth{0pc} 
\tabletypesize{\scriptsize}
\tablecaption{Atomic data used in this work \label{t2}} 
\tablehead{\colhead{Ions}    &   \colhead{$\lambda_{\rm lab}$} &   \colhead{$f$} & \colhead{Note} \\ 
\colhead{}  &\colhead{(\AA)} & \colhead{} &\colhead{}}
\startdata
\ion{C}{2}  	 & 1036.337  &   $1.18 \times 10^{-1}$        &  saturated ($a$)	\\
\ion{C}{2*}  	 & 1037.018  &   $1.18 \times 10^{-1}$        &  	\\
\ion{C}{3}  	 & 977.020   &   $7.58 \times 10^{-1}$        & blended ($b$) 	\\
\ion{N}{1}  	 & 952.303   &   $1.70 \times 10^{-3}$        &  	\\
		 & 952.415   &   $1.87 \times 10^{-3}$        &  	\\
		 & 953.415   &   $1.32 \times 10^{-2}$        &  	\\
	       	 & 953.655   &   $2.50 \times 10^{-2}$	      &  	\\
		 & 953.970   &   $3.48 \times 10^{-2}$	      &  	\\
	         & 963.990   &   $1.48 \times 10^{-2}$	      &  	\\
	         & 964.626   &   $9.43 \times 10^{-3}$	      &  	\\
	         & 965.041   &   $4.02 \times 10^{-3}$        &  	\\
	         & 1134.165  &   $1.52 \times 10^{-2}$        &  airglow ($c$)	\\
	         & 1134.415  &   $2.97 \times 10^{-2}$        &  airglow ($c$)	\\	
	         & 1134.980  &   $4.35 \times 10^{-2}$        &  airglow ($c$)	\\	
\ion{N}{2}  	 & 1083.994  &   $1.15 \times 10^{-1}$        & saturated ($a$) \\
\ion{O}{1}  	 & 919.658   &   $9.48 \times 10^{-4}$        &  	\\
		 & 919.917   &   $1.78 \times 10^{-4}$        &  	\\
		 & 921.857   &   $1.19 \times 10^{-3}$        &  	\\
		 & 922.200   &   $2.45 \times 10^{-4}$        &  	\\
		 & 924.950   &   $1.54 \times 10^{-3}$        &  	\\
		 & 925.446   &   $3.50 \times 10^{-4}$        &  	\\
		 & 929.517   &   $2.29 \times 10^{-3}$        &  	\\	     
		 & 930.257   &   $5.37 \times 10^{-4}$        &  	\\	    
		 & 936.630   &   $3.65 \times 10^{-3}$        &  	\\	    
		 & 948.686   &   $6.31 \times 10^{-3}$	      &  	\\
		 & 950.885   &   $1.58 \times 10^{-3}$	      &  	\\		     
	         & 976.448   &   $3.31 \times 10^{-3}$	      &  	\\
	         & 1039.230  &   $9.20 \times 10^{-3}$	      & airglow ($c$) 	\\
\ion{Si}{2}  	 & 1020.699  &   $1.64 \times 10^{-2}$        &  	\\
\ion{Ar}{1}  	 & 1048.220  &   $2.63 \times 10^{-1}$        &  	\\
	  	 & 1066.660  &   $6.75 \times 10^{-2}$        & blended ($d$) 	\\
\ion{P}{2}  	 & 1152.818  &   $2.45 \times 10^{-1}$        &  	\\
\ion{Fe}{2}  	 & 1055.262  &   $7.50 \times 10^{-3}$        &  	\\
		 & 1063.176  &   $5.47 \times 10^{-2}$        &  	\\
		 & 1096.877  &   $3.20 \times 10^{-2}$        &  	\\
		 & 1125.448  &   $1.60 \times 10^{-2}$        &  	\\
		 & 1143.226  &   $1.77 \times 10^{-2}$        &  	\\
		 & 1144.938  &   $1.06 \times 10^{-1}$        &  	\\
\ion{Fe}{3}  	 & 1122.524  &   $5.44 \times 10^{-2}$        &  	\\
\enddata
\tablecomments{Rest frame vacuum wavelengths and oscillator strengths are from Morton 
(private communication, 2000), except for the $f$-values of \ion{Fe}{2} which are from \citet{howk00}. 
($a$) These lines are generally strong and usually only a lower limit can be estimated (except in few cases 
where the column densities are extremely small). 
($b$) Interstellar \ion{C}{3} can be blended with the stellar component if the star contains some metals. 
($c$) Those lines can be affected by airglow emission lines, especially the \ion{N}{1}
triplet. 
($d$) \ion{Ar}{1} $\lambda$1066 can be blended with stellar \ion{Si}{4} $\lambda$1066
if the star contains some metals. 
}
\end{deluxetable}

\begin{deluxetable}{lcc}
\tablecolumns{3}
\tablewidth{0pc} 
\tabletypesize{\footnotesize}
\tablecaption{Comparison of \ion{O}{1} and \ion{N}{1} $b$-values \label{t3}} 
\tablehead{\colhead{Sightline}    & \colhead{$b$ [\ion{O}{1}]}    & \colhead{$b$ [\ion{N}{1}]}  \\
\colhead{}    &   \colhead{(\km)}  &   \colhead{(\km)}}
\startdata
WD\,0004$+$330 &    $ 6.3 \pm \, ^{0.3}_{0.3} $ &  $ 4.8 \pm \, ^{0.4}_{0.3} $   \\
WD\,0050$-$332 &    $ 5.3 \pm \, ^{0.4}_{0.6} $ &  $ 5.3: $   \\
WD\,0131$-$163 &    $ 5.5 \pm \, ^{0.1}_{0.2} $ &  $ 5.5: $   \\
WD\,0455$-$282 &    $ 5.8 \pm \, ^{0.5}_{0.3} $ &   \nodata   \\
WD\,0715$-$703 &    $ 6.4 \pm \, ^{0.4}_{0.5} $ &  $ 6.2 \pm \, ^{0.5}_{0.6} $   \\
WD\,1017$-$138 &    $ 5.7 \pm \, ^{0.9}_{1.0} $ &  $ 6.2 \pm \, ^{1.1}_{0.5} $   \\
WD\,1202$+$608 &    $ 5.4 \pm \, ^{0.3}_{0.2} $ &  $ 4.8 \pm \, ^{0.3}_{0.2} $   \\
WD\,1234$+$481 &    $ 5.0 \pm \, ^{0.4}_{0.3} $ &  $ 5.0: $   \\
WD\,1528$+$487 &    $ 5.8 \pm \, ^{0.4}_{0.3} $ &  $ 6.3 \pm \, ^{0.3}_{0.2} $   \\
WD\,1615$-$154 &    $ 4.3: $ 		     &  $ 4.3 \pm \, ^{0.3}_{0.2} $   \\
WD\,1631$+$781 &    $ 6.4 \pm \, ^{0.3}_{0.4} $ &  $ 4.7 \pm \, ^{0.3}_{0.2} $   \\
WD\,1636$+$351 &    $ 6.1 \pm \, ^{2.2}_{1.5} $ &  $ 5.7 \pm \, ^{1.2}_{1.0} $   \\
WD\,1800$+$685 &    $ 8.4 \pm \, ^{0.5}_{0.6} $ &  $ 8.8 \pm \, ^{0.3}_{0.2} $   \\
WD\,1844$-$223 &    $ 4.5 \pm \, ^{0.4}_{0.3} $ &  $ 4.3 \pm \, ^{0.8}_{0.7} $   \\
WD\,2004$-$605 &    $ 4.3 \pm \, ^{0.2}_{0.3} $ &  $ 4.5 \pm \, ^{0.4}_{0.3} $   \\
WD\,2111$+$498 &    $ 8.0 \pm \, ^{1.6}_{1.0} $ &  $ 7.9 \pm \, ^{1.5}_{1.0} $   \\
WD\,2124$-$224 &    $ 10.7 \pm \, ^{0.2}_{0.3} $&  $ 10.7 \pm \, ^{1.0}_{0.9} $  \\ 
WD\,2152$-$548 &    $ 5.0 \pm \, ^{0.3}_{0.4} $ &  $ 5.4 \pm \, ^{0.8}_{0.6} $   \\
WD\,2331$-$475 &    $ 8.6 \pm \, ^{0.3}_{0.4} $ &  $ 7.8 \pm \, ^{1.2}_{0.7} $   \\
\enddata
\tablecomments{``$:$'' means that the $b$-value was taken from the other species; see text.}
\end{deluxetable}

\begin{deluxetable}{lccccccccccc}
\tablecolumns{12}
\tablewidth{0pt}
\tabletypesize{\scriptsize}
\tablecaption{Column density summary \label{t4}} 
\tablehead{\colhead{Star}    &   \colhead{\ion{C}{2}}&   \colhead{\ion{C}{2*}}&   \colhead{\ion{C}{3}}&\colhead{\ion{O}{1}}&  
\colhead{\ion{N}{1}} &  \colhead{\ion{N}{2}} &
\colhead{\ion{Si}{2}}    &   \colhead{\ion{Ar}{1}}&   \colhead{\ion{Fe}{2}}  &  \colhead{\ion{Fe}{3}}&  
\colhead{\ion{P}{2}}}
\startdata
	    	& 	     & 	       &         & 16.20   & 15.15  &          &          &	    & 13.55   &           & 12.83    \\
WD\,0004$+$330  &$>$14.38    & $>$13.51& $>$13.40& 16.35   & 15.30  & $>$14.12 &$>$14.34  &$>$13.80 & 13.60   & $<$12.95  & 12.85    \\
	    	& 	     & 	       &         & 16.50   & 15.45  &          &          &  	    & 13.68   &           & 12.87    \\
\\
	    	& 	     & 12.29   & 	 & 15.21   & 14.01   &          &	  &	   & 12.80   &         &	  \\
WD\,0050$-$332  &$>$14.09    & 12.55   &$<$13.14 & 15.26   & 14.30   & $>$13.91 &$<$13.60 &$<$12.67& 13.06   & \nodata &$<$12.39  \\
	    	& 	     & 12.71   & 	 & 15.32   & 14.60   &          &	  &	   & 13.22   &         &	  \\
\\
	    	& 	     & 12.84   & 	 & 15.83   & 14.80   &          & 13.80   &12.90   & 13.51   &         &12.66	  \\
WD\,0131$-$163  &$>$14.20    & 13.00   &$<$13.27 & 15.89   & 14.91   & $>$14.11 & 13.93   &12.97   & 13.54   & \nodata &12.78	  \\
	    	& 	     & 13.12   & 	 & 15.97   & 15.04   &          & 14.05   &13.03   & 13.57   &         &12.87	  \\
\\
	    	&13.54       &         & 	 &         &         & 	        &	  &	   &	     &         &	  \\
WD\,0455$-$282a &13.58       &$<$12.68 &$<$12.66 &$<$14.12 & $<$13.13& $<$13.47 &$<$12.88 &$<$12.62& $<$12.73& \nodata & $<$12.39 \\
	    	&13.62       &         & 	 &         &         & 	        &	  &	   &	     &         &	  \\
\\
	    	& 	     & 12.46   &         &         & 13.29   &          &         &	   & 12.83   &         &	  \\
WD\,0455$-$282b &$>$14.20    & 12.68   &$>$13.72 & 14.91:  & 13.44   & $>$13.92 & 13.34:  &$<$12.62& 12.94   &$<$12.99 &$<$12.39  \\
	    	& 	     & 12.83   &         &         & 13.59   &          &         &	   & 13.02   &         &	  \\
\\
	    	& 	     & 13.23   & 	 & 14.82   &  13.83  & 	        &13.51    &	   & 13.03   &         &	  \\
WD\,0501$+$527  &$>$15.39    & 13.26   & \nodata & 14.86   &  13.87  & $>$13.83 &13.52    &$<$12.00& 13.05   & \nodata & $<$11.85 \\
	    	& 	     & 13.28   & 	 & 14.90   &  13.91  & 	        &13.53    &	   & 13.07   &         &	  \\
\\
	    	& 13.79     &          & 12.59   & 14.18   &	     & 13.43    &12.75    &	   &	     &         &	  \\
WD\,0549$+$158  & 13.81     & $<$12.84 & 12.72   & 14.27   & $<$13.60& 13.49    &12.99    &$<$12.46& $<$12.64& $<$13.12& $<$12.23  \\
	    	& 13.84     &          & 12.81   & 14.34   &	     & 13.54    &13.14    &	   &	     &         &	  \\
\\
	    	& 	    & 13.04   & 	 & 15.22   & 14.27   &          &	  &12.79   & 12.79   &         &	  \\
WD\,0621$-$376  &$>$14.19   & 13.09   &$<$13.31  & 15.26   & 14.34   & $>$13.80 &$<$13.68 &12.86   & 12.94   & \nodata & \nodata  \\
	    	& 	    & 13.14   & 	 & 15.31   & 14.41   &          &	  &12.93   & 13.05   &         &	  \\
\\
	    	& 	    & 13.31   &          & 15.81   & 14.89   &          & 13.97   &13.09   & 13.52   &         & 11.76    \\
WD\,0715$-$703  &$>$14.29   & 13.42   & $>$13.22 & 15.90   & 14.96   & $>$14.00 & 14.09   &13.17   & 13.56   & $<$13.34& 12.25    \\
	    	& 	    & 13.51   &          & 16.00   & 15.04   &          & 14.18   &13.25   & 13.61   &         & 12.47    \\
\\
	    	& 	    & 12.93   & 13.05   & 15.73   & 14.46    &          & 13.90   &13.02   & 13.55   &         & 12.18    \\
WD\,1017$-$138  &$>$14.10   & 13.10   & 13.14   & 15.95   & 14.55    & $>$13.90 & 14.06   &13.17   & 13.60   & \nodata & 12.23    \\
	    	& 	    & 13.23   & 13.21   & 16.40   & 14.64    &          & 14.17   &13.28   & 13.65   &         & 12.57    \\
\\
	    	& 	    & 12.28   & 	& 14.57   & 13.12   &	        & 14.05   &11.88   & 13.68   &	       & 	  \\
WD\,1202$+$608a &$>$14.83   & 12.60   &$>$14.14	& 14.70   & 13.37   & $>$14.38  & 14.11   &12.18   & 13.72   & \nodata & \nodata  \\
	    	& 	    & 12.78   & 	& 14.84   & 13.42   &	        & 14.17   &12.55   & 13.77   &	       & 	  \\
\\
	    	& 	    & 12.94   &	        & 15.63   & 14.75   &           & 14.13   &12.90   & 13.70   &         &	  \\
WD\,1202$+$608b &$>$14.65   & 13.07   &$>$14.14 & 15.65   & 14.77   & $>$14.10  & 14.19   &13.07   & 13.74   & \nodata & \nodata  \\
	    	& 	    & 13.17   &	        & 15.68   & 14.79   &           & 14.25   &13.19   & 13.78   &         &	  \\
\\
	    	& 	    & 13.07   &	        & 15.69   & 14.73   & 	        & 14.32   &13.13    & 13.50  &           & 12.50  \\
WD\,1211$+$332  &$>$14.39   & 13.14   &$>$13.49 & 15.74   & 14.77   &$>$14.03   & 14.36   &13.16    & 13.54  & $<$13.43  & 12.57  \\
	    	& 	    & 13.19   &	        & 15.79   & 14.81   & 	        & 14.40   &13.19    & 13.58  &           & 12.63  \\
\\
	    	&	    & 12.74   &	        & 15.56   & 14.58   &          & 13.65    &12.83    & 12.92   &         & 12.14    \\
WD\,1234$+$481  &$>$14.23   & 12.89   &$>$13.60 & 15.63   & 14.62   & $>$14.15 & 13.87    &12.90    & 13.07   & \nodata & 12.48    \\
	    	& 	    & 13.01   &	        & 15.71   & 14.67   &          & 14.01    &12.96    & 13.19   &         & 12.66    \\
\\
	    	& 13.62  &            &     	& 14.26   & 13.25   & 12.96    &	  &11.71    &	       &         &         \\
WD\,1254$+$223  & 13.64  & $<$12.66   & 12.20:  & 14.38   & 13.30   & 13.07    &$<$12.69  &12.04    & $<$12.54 & $<$12.94&$<$12.30 \\
	    	& 13.66  &            &  	& 14.58   & 13.45   & 13.16    &	  &12.23    &	       &         &         \\
\\
	    	& 14.66  &           & 12.54    & 14.48   & 13.48   & 13.54    &12.74     &11.71    & 12.15    &         &          \\
WD\,1314$+$293  & 14.83  & $<$12.73   & 12.59    & 14.51   & 13.51   & 13.62    &12.85     &11.96    & 12.17    & $<$12.64&$<$11.97  \\
	    	& 14.92  &           & 12.64    & 14.54   & 13.54   & 13.69    &12.94     &12.11    & 12.19    &         &          \\
\\
	    	&13.26   &           &	        &	   &	     &          &	  &	    &	       &         &           \\
WD\,1528$+$487a &13.32   &$<$12.88   &$>$13.55: &$<$14.07  &$<$13.46 & $<$13.22 &$<$13.04 &$<$12.55 & $<$13.22 & $<$13.30&$<$12.56   \\
	    	&13.36   &           &	        &	   &	     &          &	  &	    &	       &         &           \\
\\
	    	& 	 & 12.89     &          & 15.72   & 14.95   &	       & 13.80   &13.42   & 13.24   &		& 12.10   \\
WD\,1528$+$487b &$>$14.23& 12.99     &$>$13.55: & 15.80   & 14.99   & $>$14.03 & 13.91   &13.46   & 13.36   & $<$13.30  & 12.40    \\
	    	& 	 & 13.05     &	        & 15.88   & 15.04   &	       & 14.01   &13.49   & 13.45   &		& 12.70    \\
\\
	    	& 	 & 13.08     & 12.78    & 15.68   & 14.93   &13.61     &	 &13.27   & 13.44   &	        & 12.39	 \\
WD\,1615$-$154  &$>$14.03& 13.17     & 12.96    & 15.78   & 15.03   &13.71     & $<$13.94&13.29   & 13.50   & $<$13.16  & 12.45    \\
	    	& 	 & 13.25     & 13.09    & 15.89   & 15.18   &13.79     &	 &13.31   & 13.57   &	        & 12.59	 \\
\\
	    	& 	 & 13.02     &	        & 15.81   & 15.18   &	       & 14.01   &13.45   & 13.20   &          & 12.59    \\
WD\,1631$+$781  &$>$14.32& 13.17     &$>$13.41  & 15.90   & 15.28   &$>$14.12  & 14.11   &13.53   & 13.23   & $<$12.68 & 12.71    \\
	    	& 	 & 13.32     &	        & 15.99   & 15.38   &	       & 14.21   &13.61   & 13.26   &          & 12.83    \\
\\
	    	&15.65   &           &    	& 15.48   & 14.58   &14.72     & 13.59   &13.20   & 13.01   &         & 12.01    \\
WD\,1634$-$573  &15.90   & \nodata   & $<$15.4  & 15.51   & 14.62   &14.90     & 13.76   &13.26   & 13.08   & \nodata & 12.08    \\
	    	&16.25   &           &    	& 15.54   & 14.66   &15.08     & 13.89   &13.32   & 13.15   &         & 12.15    \\
\\
	    	& 	 & 13.23     &	 	& 15.58	  &         &          & 14.15   &13.64   & 13.94   &         & 12.70    \\
WD\,1636$+$351  &$>$14.25& 13.34     &$>$13.27  & 15.71   &$>$15.08 &$>$14.08  & 14.30   &13.70   & 13.98   & $<$13.41& 12.75    \\
	    	& 	 & 13.44     &	 	& 15.95   &         &          & 14.42   &13.76   & 14.02   &         & 12.80    \\
\\
		& 	 & 13.57     &	        & 16.02   & 15.02   &          & 14.44   &13.81   & 13.87   &  12.14  & 12.80    \\
WD\,1800$+$685	&$>$14.47& 13.63     &$>$13.56  & 16.12   & 15.05   & $>$14.21 & 14.53   &13.83   & 13.97   &  12.80  & 12.94    \\
		& 	 & 13.67     &	        & 16.26   & 15.09   &          & 14.60   &13.93   & 14.07   &  13.05  & 13.05    \\
\\
	    	& 	 & 12.87     &	 	& 15.85   & 14.91   &          & 14.28   &13.54   & 13.42   &         & 12.45    \\
WD\,1844$-$223  &$>$14.16& 13.04     &$>$13.25	& 15.97   & 15.06   & $>$13.82 & 14.38   &13.58   & 13.53   & $<$13.11& 12.59    \\
	    	& 	 & 13.16     &	 	& 16.15   & 15.36   &          & 14.45   &13.62   & 13.66   &         & 12.69    \\
\\
	    	&        & 12.81   & 13.12	& 15.57   & 14.76   &          & 13.92   &13.18   & 13.18   &         & 12.30    \\
WD\,2004$-$605  &$>$14.02& 12.99   & 13.15	& 15.65   & 14.84   & $>$13.85 & 14.07   &13.26   & 13.25   &$<$12.90 & 12.37    \\
	    	& 	 & 13.17   & 13.17	& 15.73   & 14.92   &          & 14.22   &13.34   & 13.32   &         & 12.44    \\
\\
	    	&        & 13.35   &         	& 16.01   & 15.20   &          & 14.16   &13.40   & 13.69   &         & 12.39    \\
WD\,2011$+$395  &$>$14.39& 13.38   & $<$13.48 	& 16.06   & 15.48   & $>$14.34 & 14.23   &13.43   & 13.72   &$<$14.09 & 12.59    \\
	    	& 	 & 13.41   &         	& 16.12   & 15.61   &          & 14.29   &13.93   & 13.75   &         & 12.72    \\
\\
	    	&        & 13.05   &		& 15.00   & 13.97   &          & 13.31   &12.46   & 13.29   &         &	  	 \\
WD\,2111$+$498  &$>$14.12& 13.12   &$<$13.00	& 15.06   & 14.00   & $>$13.91 & 13.35   &12.58   & 13.23   & \nodata & $<$12.47 \\
	    	& 	 & 13.18   &		& 15.11   & 14.08   &          & 13.31   &12.68   & 13.35   &         &	  	 \\
\\
	    	& 	 & 13.41   &         	& 15.91   & 14.93   &          & 14.44   &13.50   & 14.18   &         & 12.62    \\
WD\,2124$-$224  &$>$14.57& 13.47   & $>$13.75  	& 15.94   & 14.97   & $>$14.32 & 14.49   &13.57   & 14.22   & \nodata & 12.74    \\
	    	& 	 & 13.52   &         	& 15.97   & 15.04   &          & 14.53   &13.64   & 14.27   &         & 12.83    \\
\\
	    	& 	 & 13.73   &	 	& 16.17   & 15.48   &          &	  &	   &	     &         &	  \\
WD\,2148$+$286  &$>$14.43& 13.75   & $>$13.54  	& 16.22   & 15.55   & $>$14.13 & \nodata  &$>$13.60& \nodata & \nodata & \nodata  \\
	    	& 	 & 13.76   &	 	& 16.27   & 15.62   &          &	  &	   &	     &         &	  \\
\\
	    	& 	 & 12.65   &	 	& 15.42   & 14.37   &          & 13.38   &12.56   & 13.34   &         &	  	  \\
WD\,2152$-$548  &$>$14.30& 12.95   &$>$13.37   	& 15.49   & 14.42   & $>$14.12 & 13.76   &12.72   & 13.37   &$<$13.23 & $<$12.84  \\
	    	& 	 & 13.13   &	 	& 15.59   & 14.48   &          & 13.97   &12.90   & 13.40   &         &	  	  \\
\\
	    	& 	 & 12.79   & 	 	& 15.31   & 14.27   & 14.45    & 13.92   &12.77   & 13.17   &         & 11.95    \\
WD\,2211$-$495  &$>$14.34& 12.88   &$>$13.58   	& 15.34   & 14.30   & 14.60    & 14.00   &12.82   & 13.25   &\nodata  & 12.05    \\
	    	& 	 & 12.96   & 	 	& 15.38   & 14.33   & 14.75    & 14.08   &12.87   & 13.33   &         & 12.15    \\
\\
	    	& 	 & 13.56   &	 	& 16.53   & 15.66   &	        &            &	         & 13.96   &         &    	 \\
WD\,2247$+$583  &$>$14.41& 13.59   &$>$13.31   	& 16.72   & 15.73   &$>$13.95   &$>14.52$    &$>13.50$   & 14.03   & $<$13.14& \nodata  \\
	    	& 	 & 13.63   &	 	& 16.87   & 15.80   &	        &	     &		 & 14.10   &         &    	  \\
\\
	    	& 	 & 13.05   &	 	& 15.63   & 14.72   &          &14.05    &13.09   & 13.25   &         & 12.24    \\
WD\,2309$+$105  &$>$14.20& 13.08   &$>$13.37   	& 15.67   & 14.75   &$>$13.81  &14.07    &13.14   & 13.30   & \nodata & 12.29    \\
	    	& 	 & 13.10   &	 	& 15.71   & 14.78   &          &14.10    &13.21   & 13.35   &         & 12.34    \\
\\
	    	& 	 & 13.18   &	 	& 15.42   & 14.48   &          & 13.89   &12.67   & 13.27   &         & 12.08    \\
WD\,2331$-$475  &$>$14.43& 13.27   &$>$13.60   	& 15.48   & 14.53   & $>$14.27 & 13.99   &12.77   & 13.32   & \nodata & 12.18   \\
	    	& 	 & 13.36   &	 	& 15.54   & 14.58   &          & 14.09   &13.87   & 13.37   &         & 12.28    \\
\enddata
\tablecomments{For a given sightline and a given species, there are 3 entries: 
the second number corresponds to the most likely value, while the first and third 
correspond to $-1\sigma$ and  $+1\sigma$, respectively. 
$<$ indicates a 3$\sigma$ upper limit detection, except in the case of 
\ion{C}{3} where the upper limit reflects that
this line may be blended with the \ion{C}{3} photospheric lines. 
$>$ indicates a lower limit, i.e. the line is saturated. Colon means that the value is uncertain.
See \S~\ref{col} and Table~\ref{t1} for more details and references, respectively.
Two entries for one sightline (denoted by a and b) indicate that two components are detected with {\em FUSE}:
for WD\,0455$-$282, component a is at $-58$ \km\ and b at $-2$ \km; 
for WD\,1202$+$608, component a is at $-58$ \km\ and b at $-2$ \km; and
for WD\,1528$+$487, component a is at $-56$ \km\ and b at $-16$ \km.
}
\end{deluxetable}

\begin{deluxetable}{lccccccc}
\tablecolumns{8}
\tablewidth{0pc} 
\tabletypesize{\scriptsize}
\tablecaption{H$_2$ measurements \label{t5}} 
\tablehead{\colhead{Sightline}    & \colhead{$N_{J=0}$}    & \colhead{$N_{J=1}$} & \colhead{$N_{J=2}$} & \colhead{$N_{J=3}$} & \colhead{$N_{\rm all}$} & \colhead{$b$} & \colhead{$T_{\rm ex}$} \\
\colhead{}    &  \colhead{}    &  \colhead{}    &  \colhead{}    &  \colhead{}   &  \colhead{}    &   \colhead{(\km)}   &   \colhead{(K)}}
\startdata
WD\,0004$+$330&   $13.45 \pm 0.06$ (1)		 & $14.35 \pm 0.04 $ (22)   		& $13.71 \pm 0.04 $ (10)    		& $13.52 \pm 0.07 $ (4)		    	& 14.46  &	    $  3.8 \pm 0.7$ &$ 314 \pm 42 $$^a$\\
WD\,1636$+$351&   $14.21 \pm 0.20$ (7)     	 & $14.83 \pm 0.07 $ (25)   		& $14.30 \pm 0.15  $ (13)   		& $14.04 \pm \,^{0.10}_{0.07}$ (3)  	& 15.05  &	    $  3.1 \pm 0.2  $  & $314 \pm 31$ \\
WD\,1800$+$685&   $14.02\pm\,^{0.26}_{0.10} $ (5)& $14.49 \pm 0.15 $ (12)  		& $14.00 \pm 0.10 $ (12)		&$13.79 \pm 0.18 $ (1)	 	    	& 14.75  &	    $  5.0 \pm \,^{4.2}_{2.1}$  	    &$ 301 \pm 38 $    \\
WD\,2011$+$395&   \nodata      			 & $14.55 \pm 0.10 $ (13)  		& $14.05 \pm 0.20 $ (2)			&$13.83 \pm 0.20 $ (2)	  	    	& 14.22  &	    $  2.8 \pm \,^{0.4}_{0.3}$      &  $349 \pm 74$	       \\
WD\,2148$+$286&$14.32 \pm\,^{0.11}_{0.09}$ (4) 	 & $15.25 \pm\,^{0.24}_{0.12}$ (8) 	& $14.34 \pm \,^{0.15}_{0.16}$ (13)	& $14.18 \pm \,^{0.14}_{0.15}$ (14) 	& 15.38  &	    $	3.0 \pm 0.6 $ &  $297 \pm 35 $\\
WD\,2247$+$583&   $14.24 \pm 0.03	 $ (6)   & $14.94 \pm 0.04$ (13)		& $14.23 \pm 0.03$   (14)      		& $13.94 \pm 0.04 $	(7)  	    	& 15.11  &	    $ 4.1 \pm 0.4   $  &$ 276 \pm 8 $ \\
\enddata
\tablecomments{The number of lines used to measure the column density is shown in parentheses. 
For WD\,0004$+$330, WD\,2148$+$286, and WD\,2247$+$583, the column densities and $b$-values 
are from Oliveira et al. (2003, private communication) Sonneborn et al. (2003, erratum in preparation), and 
\citet{oliveira02}, respectively. $a:$ For WD\,0004$+$330, only the $J$-levels $1,2,3$ could be 
fitted in the temperature relation within 1$\sigma$ error bar, $T_{01} \approx 1363$ K.
}
\end{deluxetable}

\newpage
\begin{deluxetable}{lccc}
\tablecolumns{4}
\tablewidth{0pc} 
\tabletypesize{\footnotesize}
\tablecaption{[\ion{Ar}{1}/\ion{O}{1}],  [\ion{N}{1}/\ion{O}{1}], and \ion{N}{1}/(\ion{N}{2}\,+\,\ion{N}{1}) ratios \label{t6}} 
\tablehead{\colhead{Sightlines}    & \colhead{[\ion{Ar}{1}/\ion{O}{1}]}     & \colhead{[\ion{N}{1}/\ion{O}{1}]} & \colhead{\ion{N}{1}/(\ion{N}{2}+\ion{N}{1})}}
\startdata
 WD\,0004$+$330  &$ >-0.41 		       $  & $ -0.29 \pm\,_{0.23 }^{0.20}$  & $< 0.94 $  \\
 WD\,0050$-$332  &$ <-0.45		       $  & $ -0.20  \pm\,_{0.30}^{0.30}$  & $ <0.71 $  \\
 WD\,0131$-$163  &$  -0.78 \pm\,_{0.10 }^{0.10}$  & $ -0.22  \pm\,_{0.13}^{0.15}$  & $ <0.86 $  \\
 WD\,0455$-$282b &$ <-0.15		       $  & $ -0.71:			$  & $ <0.25 $  \\
 WD\,0501$+$527  &$ <-0.72		       $  & $ -0.23  \pm\,_{0.06}^{0.06}$  & $ <0.52 $  \\
 WD\,0549$+$158  &$ < 0.33		       $  & $< 0.09			$  & $ 0.56: $  \\
 WD\,0621$-$376  &$  -0.26  \pm\,_{0.08}^{0.08}$  & $ -0.16  \pm\,_{0.08}^{0.08}$  & $ <0.77 $  \\
 WD\,0715$-$703  &$  -0.59  \pm\,_{0.13}^{0.12}$  & $ -0.18  \pm\,_{0.12}^{0.12}$  & $ <0.90 $  \\
 WD\,1017$-$138  &$  -0.64  \pm\,_{0.30}^{0.45}$  & $ -0.64  \pm\,_{0.25}^{0.45}$  & $ <0.82 $  \\
 WD\,1202$+$608a &$  -0.38  \pm\,_{0.36}^{0.37}$  & $ -0.57  \pm\,_{0.31}^{0.16}$  & $ <0.09 $  \\
 WD\,1202$+$608b &$  -0.44  \pm\,_{0.17}^{0.12}$  & $ -0.12  \pm\,_{0.03}^{0.04}$  & $ <0.82 $  \\
 WD\,1211$+$332  &$  -0.44  \pm\,_{0.05}^{0.06}$  & $ -0.21  \pm\,_{0.07}^{0.06}$  & $ <0.85 $  \\
 WD\,1234$+$481  &$  -0.59  \pm\,_{0.10}^{0.10}$  & $ -0.25  \pm\,_{0.08}^{0.09}$  & $ <0.75 $  \\
 WD\,1254$+$223  &$  -0.20  \pm\,_{0.38}^{0.26}$  & $ -0.32  \pm\,_{0.13}^{0.30}$  & $ 0.63\pm\,_{0.02}^{0.10} $  \\
 WD\,1314$+$293  &$  -0.41  \pm\,_{0.25}^{0.15}$  & $ -0.24  \pm\,_{0.04}^{0.04}$  & $ 0.44\pm\,_{0.03}^{0.03} $  \\
 WD\,1528$+$487b &$  -0.20  \pm\,_{0.11}^{0.07}$  & $ -0.05  \pm\,_{0.11}^{0.08}$  & $ <0.90 $  \\
 WD\,1615$-$154  &$  -0.35  \pm\,_{0.10}^{0.11}$  & $  0.01  \pm\,_{0.15}^{0.18}$  & $ 0.95\pm\,_{0.04}^{0.08} $  \\
 WD\,1631$+$781  &$  -0.23  \pm\,_{0.13}^{0.12}$  & $  0.14  \pm\,_{0.15}^{0.13}$  & $ <0.94 $  \\
 WD\,1634$-$573  &$  -0.11  \pm\,_{0.07}^{0.07}$  & $ -0.13  \pm\,_{0.05}^{0.05}$  & $ 0.34\pm\,_{0.09}^{0.13} $  \\
 WD\,1636$+$351  &$   0.13  \pm\,_{0.15}^{0.24}$  & $> 0.13			$  & $ <0.91 $  \\
 WD\,1800$+$685  &$  -0.15  \pm\,_{0.10}^{0.16}$  & $ -0.31  \pm\,_{0.11}^{0.14}$  & $ <0.87 $  \\
 WD\,1844$-$223  &$  -0.25  \pm\,_{0.13}^{0.18}$  & $ -0.15  \pm\,_{0.21}^{0.33}$  & $ <0.94 $  \\
 WD\,2004$-$605  &$  -0.25  \pm\,_{0.12}^{0.11}$  & $ -0.05  \pm\,_{0.12}^{0.11}$  & $ <0.91 $  \\
 WD\,2011$+$395  &$  -0.48  \pm\,_{0.06}^{0.50}$  & $  0.19  \pm\,_{0.29}^{0.17}$  & $ <0.93 $  \\
 WD\,2111$+$498  &$  -0.34  \pm\,_{0.14}^{0.11}$  & $ -0.30  \pm\,_{0.07}^{0.09}$  & $ <0.55 $  \\
 WD\,2124$-$224  &$  -0.23  \pm\,_{0.04}^{0.08}$  & $ -0.20  \pm\,_{0.05}^{0.08}$  & $ <0.82 $  \\
 WD\,2148$+$286  &$ >-0.48		       $  & $  0.09  \pm\,_{0.09}^{0.08}$  & $ <0.96 $  \\
 WD\,2152$-$548  &$  -0.63  \pm\,_{0.18}^{0.20}$  & $ -0.31  \pm\,_{0.09}^{0.11}$  & $ <0.67 $  \\
 WD\,2211$-$495  &$  -0.38  \pm\,_{0.07}^{0.06}$  & $ -0.28  \pm\,_{0.05}^{0.05}$  & $  0.33\pm\,_{0.06}^{0.08} $  \\
 WD\,2247$+$583  &$  -0.65  \pm\,_{0.19}^{0.21}$  & $ -0.23  \pm\,_{0.21}^{0.16}$  & $ <0.98 $  \\
 WD\,2309$+$105  &$  -0.39  \pm\,_{0.07}^{0.10}$  & $ -0.16  \pm\,_{0.05}^{0.05}$  & $ <0.89 $  \\
 WD\,2331$-$475  &$  -0.64  \pm\,_{0.10}^{0.10}$  & $ -0.31  \pm\,_{0.06}^{0.07}$  & $ <0.63 $  \\
\enddata
\tablecomments{Suffix ``a'' after star names means it is a high velocity component (see \S~\ref{hvc}), 
while ``b'' means that it is the low velocity component. Colon means that the value is uncertain.
}
\end{deluxetable}

\begin{deluxetable}{lccc}
\tablecolumns{4}
\tablewidth{0pc} 
\tabletypesize{\footnotesize}
\tablecaption{\ion{C}{3}/\ion{C}{2} ratio \label{t7}} 
\tablehead{\colhead{Targets}    &   \colhead{$\log [N($\ion{C}{3}$/N($\ion{C}{2})]}
&   \colhead{$\log [N($\ion{C}{3}$/N($\ion{C}{2})]$_{\rm c}$} & \colhead{Note}}
\startdata
WD\,0050$-$332    &   $< -0.95$  &  $<-1.26$ to $<-0.95$  & \ion{C}{2} LL,  \ion{C}{3} SC?	\\
WD\,0131$-$163    &   $< -0.93$  &  $< -0.93$		  & \ion{C}{2} LL 			\\
WD\,0455$-$282a   &   $< -0.92$  &   \nodata		  & \ion{C}{3} UL		 	\\
WD\,0549$+$158    &   $ -1.09$   & $<-1.09$ to $<-1.05$   & \ion{C}{3} SC?			\\
WD\,0621$-$376    &   $< -0.90$  &  $<-1.05$ 		  & \ion{C}{2} LL,  \ion{C}{3} SC 	\\
WD\,1017$-$138    &   $<-0.96$   &   $<-0.96$		  & \ion{C}{2} LL			\\
WD\,1254$+$223    &   $-1.44:$   &  $<-1.44:$\		  &  \ion{C}{3} SC?			\\
WD\,1314$+$293    &   $-2.24$    &  $-2.39$ to $-2.24$    &  \ion{C}{3} SC?			\\
WD\,1528$+$487a   &   $> 0.23:$  &  \nodata		  &  \ion{C}{3} LL			\\
WD\,1615$-$154    &   $< -1.07$  & $< -1.07$		  &  \ion{C}{2} LL			\\
WD\,1634$-$573    &   $ <-0.50$  &  $-0.63$		  &  \ion{C}{3} SC			\\
WD\,2004$-$605    &   $<-0.87$   & $<-0.87$		  &  \ion{C}{2} LL			\\
WD\,2011$+$395    &   $< -0.91$  &   $<-1.44$  		  &  \ion{C}{2} LL,  \ion{C}{3} SC	\\
WD\,2111$+$498    &   $< -1.12$  &   $<-1.52$  		  &  \ion{C}{2} LL,  \ion{C}{3} SC	\\
\enddata
\tablecomments{The second column presents the raw measurement, while the third one includes
possible stellar correction for \ion{C}{3}. When two numbers are indicated in 
the third column, it means that only an upper limit was derived for the amount 
of photospheric \ion{C}{3} $\lambda$977. LL: lower limit. UL: upper limit.
SC: \ion{C}{3} stellar contamination. SC?: possible \ion{C}{3} stellar contamination. 
Suffix ``a'' after star names means the ratio is for a high velocity component (see \S~\ref{hvc}),
and this component is not blended with any stellar lines. 
Colon means that the value is uncertain.}
\end{deluxetable}

\begin{deluxetable}{lc}
\tablecolumns{2}
\tablewidth{0pc} 
\tabletypesize{\footnotesize}
\tablecaption{\ion{Fe}{3}/\ion{Fe}{2} ratio \label{t8}} 
\tablehead{\colhead{Targets}    &   \colhead{$\log [N($\ion{Fe}{3}$/N($\ion{Fe}{2})]}  }
\startdata
WD\,0004$+$330    &   $<  -0.65$	\\
WD\,0455$-$282b   &   $< 0.05$		\\
WD\,0715$-$703    &   $ <-0.22$		\\
WD\,1211$+$332    &   $<-0.11$		\\
WD\,1314$+$293    &   $< 0.47$		\\
WD\,1528$+$487b   &   $<-0.06$		\\
WD\,1615$-$154    &   $< -0.34$		\\
WD\,1631$+$781    &   $< -0.63$		\\
WD\,1636$+$351    &   $<-0.57$		\\
WD\,1800$+$685    &   $-1.17: $  	\\
WD\,1844$-$223    &   $<-0.42$  	\\
WD\,2004$-$605    &   $<-0.35$  	\\
WD\,2011$+$395    &   $<0.37$  		\\
WD\,2152$-$548    &   $<-0.14$  	\\
WD\,2247$+$583    &   $< -0.89 $	\\
\enddata
\tablecomments{Here ``$<$'' is a 3$\sigma$ upper limit.
Colon means that the value is uncertain.
}
\end{deluxetable}

\clearpage

\end{document}